\documentclass[fleqn,usenatbib]{mnras}

\usepackage{amsmath}
\usepackage{txfonts}

\usepackage[british]{babel} 
\usepackage[T1]{fontenc}
\usepackage{ae,aecompl}

\usepackage{graphicx}	% Including figure files
\usepackage{xspace}
\usepackage{dcolumn}
\usepackage{printlen}
\graphicspath{{./plots/}}
\newcommand{\galform}{{\sc{galform}}\xspace}
\newcommand{\grasil}{{\sc{grasil}}\xspace}

\title[The evolution of the UV-to-mm EBL]{The evolution of the UV-to-mm extragalactic background light: evidence for a top-heavy initial mass function?}

\author[W. I. Cowley et al.]{
William I. Cowley$^{1}$\thanks{E-mail: cowley@astro.rug.nl (WIC)},
Cedric G. Lacey$^{2}$,  
Carlton M. Baugh$^{2}$,
Shaun Cole$^{2}$, \newauthor
Carlos S. Frenk$^{2}$ and 
Claudia del P. Lagos$^{3\mathrm{,}4\mathrm{,}5}$  
\\
$^{1}$Kapteyn Astronomical Institute, University of Groningen, PO Box 800, NL-9700 AV Groningen, the Netherlands\\
$^{2}$Institute for Computational Cosmology, Department of Physics, University of Durham, South Road, Durham, DH1 3LE, UK\\
$^{3}$International Centre for Radio Astronomy Research (ICRAR), M468, University of Western Australia, 35 Stirling Hwy, Crawley, WA 6009,\\ Australia\\
$^{4}$Australian Research Council Centre of Excellence for All-sky Astrophysics (ASTRO 3D)\\
$^{5}$Cosmic Dawn Center (DAWN), Niels Bohr Institute, University of Copenhagen, Juliane Maries vej 30, DK-2100 Copenhagen, Denmark
}

\date{Accepted XXX. Received YYY; in original form ZZZ}

\pubyear{2017}

\begin{document}
\label{firstpage}
\pagerange{\pageref{firstpage}--\pageref{lastpage}}
\maketitle

\begin{abstract}
We present predictions for the UV-to-mm extragalactic background light (EBL) from a recent version of the \galform\ semi-analytical model of galaxy formation which invokes a top-heavy stellar initial mass function (IMF) for galaxies undergoing dynamically-triggered bursts of star formation. We combine \galform\ with the \grasil\ radiative transfer code for computing fully self-consistent UV-to-mm spectral energy distributions for each simulated galaxy, accounting for the absorption and re-emission of stellar radiation by interstellar dust. The predicted EBL is in near-perfect agreement with recent observations over the whole UV-to-mm spectrum, as is the evolution of the cosmic spectral energy distribution over the redshift range for which observations are available ($z\lesssim1$). We show that approximately 90~per~cent of the EBL is produced at $z<2$ although this shifts to higher redshifts for sub-mm wavelengths. We assess whether  the top-heavy IMF in starbursts  is necessary in order to reproduce the EBL at the same time as other key observables, and find that variant models with a universal solar-neighborhood IMF display poorer agreement with EBL observations over the whole UV-to-mm spectrum and fail to match the counts of galaxies in the sub-mm.
\end{abstract}


%%%%%%%%%%%%%%%%%%%%%%%%%%%%%%%%%%%%%%%%%%%%%%%%%%

%%%%%%%%%%%%%%%%% BODY OF PAPER %%%%%%%%%%%%%%%%%%
\section{Introduction}
The extragalactic background light (EBL) provides a record of the production of photons since (re)combination, and thus contains a wealth of information regarding various astrophysical processes over the history of the Universe. In the $0.1-1000$~$\muup$m (UV-to-mm) wavelength range it is dominated by the redshifted emission from galaxies, including the absorption and re-emission by interstellar dust of photons produced in stars. It also includes minor ($\lesssim10$~per~cent) contributions from active galactic nuclei \citep[AGN e.g.][]{Almaini99,Silva04}, intra-halo light (IHL) from diffuse halo stars no longer associated with a host galaxy \citep[e.g.][]{Zemcov:2014} and redshifted Lyman~$\alpha$ emission from the epoch of reionization \citep[e.g.][]{MitchellWynne:2015}. As such, the EBL provides strong constraints on the cosmic star formation history and on models of galaxy formation and evolution \citep[e.g.][]{Fardal:2007,Franceschini:2008,Finke:2010,Somerville:2012,Inoue:2013,Andrews:2018,Baes:2019}.

Historically, two methods have been used to observationally estimate the EBL: (i) direct detection with instruments such as the diffuse infrared background explorer \citep[DIRBE, ][]{Silverberg:1993} and the far-infrared absolute spectrophotometer \citep[FIRAS, ][]{Mather:1993} flown on the Cosmic Background Explorer satellite \citep[\emph{COBE} e.g.][]{Puget96, Fixsen98, Wright:2004}; and (ii) integrating galaxy number counts \citep[e.g.][]{MadauPozzetti:2000,Berta:2011,Bethermin12,Driver:2016}. The former requires accurate removal of foregrounds, most notably that of zodiacal light \citep[solar emission scattered by interplanetary dust e.g. ][]{Bernstein:2002, Mattila:2006} and emission from the Milky Way \citep[e.g.][]{Bernard:1994, Arendt:1998}, which have put a limit on the accuracy with which the EBL can be measured directly. The second method requires an extrapolation to faint fluxes as is discussed in more detail below.  

Integrating galaxy number counts has, until relatively recently, suffered from insufficiently deep data, particularly at far-IR wavelengths, to fully resolve the EBL. In this wavelength regime confusion noise introduced by the coarse angular resolution [$\sim20$~arcsec full width at half maximum (FWHM)] of single-dish telescopes commonly used for imaging at these wavelengths and the high surface density of detectable objects (e.g. Nguyen et al. \citeyear{Nguyen:2010}) meant that only a small fraction ($\sim15$~per~cent) of the far-IR EBL could be resolved \citep[][]{Oliver10}. The use of techniques such as gravitational lensing \citep[e.g.][]{Smail97, Knudsen08, Chen13}, stacking \citep[e.g.][]{Bethermin12, Geach13} and high resolution interferometry \citep[e.g.][]{Hatsukade13,Carniani15} has allowed galaxy number counts to be statistically estimated at fluxes fainter than the traditional confusion limit. This has resolved a much higher proportion of the EBL, and results from direct detection and from integrated number counts are now in good agreement over mid- to far-IR wavelengths. There exists a general discrepancy between integrated counts and direct measurements at optical/near-IR wavelengths however, with direct observational estimates typically being factors of $\sim2-5$ higher than those obtained from the integrated counts. This could indicate that sources of light not associated with individual galaxies (e.g. IHL) form a significant component of the EBL at these wavelengths, or that the models used in foreground removal require revision.           

Recently, a third, independent, method of estimating the EBL has shed some light on this issue. Measurements of the attenuation of high-energy (TeV) photons from blazars, which are assumed to be emitted with a well-defined power law spectrum, as they scatter with EBL photons could reveal the spectrum of the EBL. This was first illustrated by the High Energy Stereoscopic System \citep{HESS:2006}, and detailed measurements have since been performed over the full UV-to-mm range \citep[][]{BiteauWilliams:2015, MAGIC:2016}.  These independent measurements all favour the estimates from integrated number counts (though some caveats do remain e.g. their dependence on the assumed intrinsic shape of the blazar spectrum), suggesting that current zodical light models may require some revision, and that light not associated with galaxies e.g. IHL, makes a minimal contribution (see also the discussion in Driver et al. \citeyear{Driver:2016}). For this reason, throughout we take the observed EBL as being equivalent to what is obtained from integrating galaxy number counts at all UV-to-mm wavelengths.   

Here we present predictions for the EBL from the well-established semi-analytical model for galaxy formation, \galform\ \citep[e.g.][]{Cole00,Lacey16}. This provides a physical calculation of galaxy formation from high redshift ($z\gtrsim15$) to the present day ($z=0$), accounting for the main physical processes involved (e.g. gravitational collapse, gas cooling, star formation and feedback) implemented within the cold dark matter (CDM) cosmological model.  Simulated galaxy spectral energy distributions (SEDs) are computed using the radiative transfer code \grasil\ \citep{Silva98}. This means the absorption, scattering and re-emission of stellar radiation by interstellar dust is calculated completely self-consistently from the physical properties of galaxies predicted by \galform\ (e.g. gas-phase metallicity, size) and the assumed geometry and composition of the interstellar dust. The model thus provides a consistent physical framework for interpreting multi-wavelength observations over the history of the Universe.   

This combined modelling represents a significant advantage over empirical models that employ arbitrary phenomenological recipes to reproduce key observational constraints \citep[and thus forgo a physical interpretation of their predictions e.g.][]{Franceschini:2008,Dominguez:2011,Andrews:2018}, and over models that rely on empirical SED templates for calculating galaxy spectra over some (e.g. far-IR) or all of the UV-to-mm spectrum \citep[as their predicted luminosities are not necessarily internally self-consistent with the underlying galaxy formation model e.g.][]{Gilmore:2012,Somerville:2012}. Additionally, the flexibility of the semi-analytical method means that variant models in which some modelling assumptions are varied can be calculated quickly to assess their impact on reproducing various observations. This type of parameter exploration is not generally possible with the current state-of-the-art hydrodynamical cosmological galaxy formation simulations \citep[e.g.][]{Vogelsberger14, Schaye15, Nelson:2018} due to their prohibitive computational expense.          

One of the key features of the version of the \galform\ model used in this work \citep[and described fully in][]{Lacey16} is that it incorporates a top-heavy initial mass function (IMF) during periods of dynamically-triggered star formation. This feature was incorporated into the model so that it could simultaneously reproduce the number counts and redshift distribution of sub-mm galaxies observed at $850$~$\muup$m and the present day (i.e. $z=0$) optical and near-IR galaxy luminosity functions \citep{Baugh05}. The IMF used in the \cite{Lacey16} model is much less top-heavy, however, than the one implemented by Baugh et al. We investigate whether this feature is required in order for the model to reproduce the EBL at far-IR wavelengths in conjunction with other constraints such as the $K$-band luminosity function at $z=0$ and the evolution of the cosmic star formation rate density. In doing so we reassess the argument of \cite{Fardal:2007}, who suggest that these three observational datasets are incompatible with a universal IMF of the form observed in the solar neighbourhood. Fardal et al. integrated simple parametrizations of the cosmic star formation history and found that it was not possible to find histories that could reproduce the local $K$-band luminosity density and the EBL simultaneously whilst assuming a Salpeter (\citeyear{Salpeter55}) IMF (see e.g. their Figure~5). We note that \cite{Somerville:2012}, using a semi-analytical galaxy formation model assuming a universal IMF, found a reasonable match to the EBL, cosmic star formation history and present-day $K$-band luminosity function, but under-predicted the number counts of galaxies at $850 \,\mu $m. Other galaxy formation studies have considered IMF variations, such as \cite{Gargiulo:2015} and \cite{Fontanot:2017}, who invoked a top-heavy IMF in regions of high star formation in semi-analytical models, and \cite{Barber:2018}, who imposed a pressure dependent IMF in an {\tt EAGLE} hydrodynamical simulation; these studies did not consider the EBL.

This paper is structured as follows: in Section~\ref{sec:Model} we introduce the theoretical model, which incorporates a semi-analytical model of galaxy formation implemented within halo merger trees derived from a Millennium-style dark matter only $N$-body simulation \citep{Springel05,Baugh:2019} and the radiative transfer code, \grasil\ \citep{Silva98}, for computing the absorption and re-emission of stellar radiation by interstellar dust. In Section~\ref{sec:Results} we present the predictions of the model for the EBL and show how this is built up over the history of the Universe\footnote{Some of the model data presented here will be made available at \url{http://icc.dur.ac.uk/data/}.  For other requests please contact the first author.}. We also present predictions from variant models with a universal solar-neighbourhood IMF and discuss how critical this feature is for reproducing the EBL. We summarise in Section~\ref{sec:Conclusion}. Throughout we assume a flat $\Lambda$CDM cosmology with cosmological parameters consistent with recent \emph{Planck} satellite results \citep{PlanckCollaboration:2014}\footnote{$\Omega_{\rm m}=0.307$, $\Omega_{\Lambda}=0.693$, $h=0.678$, $\Omega_{\rm b}=0.0483$, $\sigma_{8}=0.829$}.
 
\section{The Model}
\label{sec:Model} 
Here we introduce our theoretical model, which combines a dark matter only $N$-body simulation, a semi-analytical model of galaxy formation (\galform) and the spectrophotometric radiative transfer code \grasil\ \citep{Silva98} for computing self-consistent UV-to-mm galaxy SEDs.
\subsection{The \emph{Planck} Millennium dark matter simulation}
Galaxies are assumed to form from baryonic condensation within the potential wells of dark matter halos, with their subsequent evolution being controlled in part by the merging history of the halo \citep{WhiteRees78}. Here halo merger trees are extracted directly from a dark matter only $N$-body simulation \citep[e.g.][]{Helly03,Jiang14}.  We use a new ($800$~Mpc)$^{3}$ Millennium-style simulation \citep{Springel05} with cosmological parameters consistent with recent \emph{Planck} satellite results \citep{PlanckCollaboration:2014}, henceforth referred to as the P--Millennium \citep[Baugh et~al. 2018; see also][]{McCullagh:2017,Cowley:2018-JWST}. The large volume, (800~Mpc)$^3$, gives the bright end of the predicted luminosity functions greater statistical precision.

The halo mass resolution of this simulation is $2.12\times10^{9}$~$h^{-1}$~M$_{\sun}$, where a halo is required to have at least $20$ dark matter particles and is defined according to the `DHalo' algorithm \citep{Jiang14}.  This mass resolution is approximately an order of magnitude better than previous dark matter simulations that were used with this galaxy formation model. For example, the MR7 simulation \citep{Springel05,Guo13} in which the \cite{Lacey16} model was originally implemented has a halo mass resolution of $1.87\times10^{10}$~$h^{-1}$~M$_{\sun}$. Not only does this mean that the model is able to make predictions for smaller mass halos (i.e. fainter galaxies) but also that the more moderate mass halos ($\sim10^{11}-10^{12}$~$h^{-1}$~M$_{\sun}$) that in this model dominate the far-IR background \citep{Cowley16} are resolved with greater precision. The P--Millennium merger trees also provide a finer temporal resolution than MR7 but this does not have any significant impact on this model.
\subsection{Semi-analytical Galaxy Formation}
Baryonic physics in \galform\ are included as a set of coupled differential equations that track the exchange of mass and metals between between the stellar, cold disc gas and hot halo gas components in a given halo.  These equations include simplified prescriptions for the physical processes (e.g. gas cooling, star formation and feedback) known to be important for galaxy formation. We discuss some of the main features of the model below and refer the interested reader to \cite{Lacey16} for more details. 

There are, however, minor changes to the values of two parameters from the model presented by Lacey et al. These are described in more detail in \cite{Baugh:2019}, see also Section~2.1.5 and Table~1 of \cite{Cowley:2018-JWST}, and mainly account for the fact that the underlying halo merger trees within which the model is implemented are generated from a dark matter simulation with improved halo mass resolution (see above) and different cosmological parameters, so that the model can reproduce the original calibration data to a similar level of fidelity. The impact that this recalibration has on the predicted EBL is discussed in Appendix~\ref{app:grasil_recal}.  
\subsubsection{Star formation and stellar initial mass function}
\label{sec:SF_IMF}
Cold disc gas is partitioned into molecular and atomic components according to the mid-plane gas pressure in the disc. The star formation rate surface density is then assumed to be proportional to the surface density of molecular gas, such that
\begin{equation}
\Sigma_{\rm SFR} = \nu_{\rm SF}\,\Sigma_{\rm mol} = \nu_{\rm SF}\,f_{\rm mol}\,\Sigma_{\rm cold}\rm,
\label{eq:star_formation_law}
\end{equation} 
where $f_{\rm mol}=R_{\rm mol}/(1 + R_{\rm mol})$, $R_{\rm mol}$ is the local ratio of molecular and atomic gas surface densities i.e. $R_{\rm mol}=\Sigma_{\rm mol}/\Sigma_{\rm atom}$ and the parameter $\nu_{\rm SF}=0.74$~Gyr$^{-1}$, based on the observations of \cite{Bigiel11}. This expression is then integrated over the whole disc to yield the global star formation rate, $\psi$. For further details of this star formation law, see  \cite{Lagos11}.  For star formation in the galactic disc a \cite{Kennicutt83} IMF is assumed.  This IMF is described by $x=0.4$ in $\mathrm{d}N/\mathrm{d}\ln m\propto m^{-x}$ for $m<1$~M$_{\sun}$ and $x=1.5$ for $m>1$~M$_{\sun}$ (for reference, a Salpeter (\citeyear{Salpeter55}) IMF has an unbroken slope of $x=1.35$).

Galaxy starbursts are triggered by dynamical processes. These are either a bar instability in the disc applying the stability criterion of (\citealt{Efstathiou:1982}, see Section 3.6.2 of Lacey et~al. for further details) or major galaxy mergers (and some gas-rich minor mergers). Throughout this work `(star)bursts' refer to such dynamically-triggered star formation event rather than, for example, to a galaxy's position on the specific star formation rate--stellar mass plane. This distinction is discussed in \cite{Cowley16SEDs}. Burst star formation takes place in a forming galactic bulge. It is assumed that $f_{\rm mol}\approx1$ in bursts and that the star formation rate depends on the dynamical timescale of the bulge as
\begin{equation}
\psi_{\rm burst} = \nu_{\rm SF,burst}M_{\rm cold,burst}\rm,
\end{equation} 
where $\nu_{\rm SF,burst}=1/\tau_{\star\rm,burst}$ and
\begin{equation}
\tau_{\star\rm,burst} = \max[f_{\rm dyn}\tau_{\rm dyn,bulge},\tau_{\rm burst,min}]\rm.
\end{equation} 
Here $\tau_{\rm dyn,bulge}$ is the dynamical timescale of the bulge and $f_{\rm dyn}$ and $\tau_{\rm burst,min}$ are model parameters.  This means that for large dynamical times the star formation rate scales mostly with the dynamical time, but has a ceiling value when the dynamical time of the bulge is short. Here $f_{\rm dyn}=20$ and $\tau_{\rm burst,min}=100$~Myr \citep{Lacey16}.  

For star formation in bursts, it is assumed that stars form with a top-heavy IMF, described by a slope of $x=1$.
 
This assumption is primarily motivated by the requirement that the model reproduce the observed far-IR/sub-mm galaxy number counts and redshift distributions \citep[e.g.][]{Baugh05,Lacey16}. It should be noted that the IMF slope in this new model is much less top-heavy than the $x=0$ one used by \cite{Baugh05}. 

The assumption of a top-heavy IMF for starburst galaxies is often seen as controversial. For example, in their review of observational studies \cite{Bastian:2010} argue against significant IMF variations in the local Universe. However, \cite{Ballero:2007} argue through chemical evolution modelling that an $x\sim1$ slope is required to explain the [Fe/H] distribution in the bulges of the Milky Way and M31. Additionally, \cite{Gunawardhana:2011} infer an IMF for nearby star-forming galaxies that becomes more top-heavy with increasing star formation rate, reaching a slope of $x\approx0.9$; and a similar IMF slope was inferred for a star-forming galaxy at $z\sim2.5$ by \cite{Finkelstein:2011}. Both of these studies use modelling of a combination of nebular emission and broadband photometry to infer the IMF slope. More recently, \cite{Romano:2017} inferred an IMF slope of $x=0.95$ in nearby starburst galaxies through modelling the observed CNO isotopic ratios. This method has since been extended to dust-obscured star-forming galaxies at $z\sim2-3$ by \cite{Zhang18}, who claim unambiguous evidence for a similarly top-heavy IMF ($x=0.95$) in four gravitationally lensed sub-mm galaxies. Evidence for a top-heavy IMF has also been found in local star-forming regions. A recent study of massive stars ($m\gtrsim15$~M$_{\sun}$) in the 30 Doradus region in the Large Magellanic Cloud found an IMF slope of $0.9\pm0.3$ \citep{Schneider:2018}\footnote{A subsequent analysis of the data used by Schneider et al. found a slightly different value, $x=1.05_{-0.14}^{+0.13}$ \citep{FarrMandel:2018}, which is still top-heavy relative to that of the solar neighbourhood and with a smaller uncertainty than determined by Schneider et al. See also \cite{Schneider:2018-comment}.}. Thus, whilst the issue of a varying IMF is far from resolved, there is a growing number of observational studies that support both our assumption and adopted value of $x=1$. We note, however, that the debate about the form of the IMF continues and a number of studies support a revision to the IMF in the opposite sense to that proposed here, in the direction of being more bottom-heavy (e.g. \citealt{Smith:2015,LaBarbera:2016,Collier:2018}). 

\subsubsection{Supernovae feedback}
The injection of energy into the ISM from supernovae (SN) is assumed to eject gas from the disc to beyond the virial radius of the halo at a rate $\dot{M}_{\rm eject}$. As SN are short-lived, this rate is proportional to the instantaneous star formation rate, $\psi$, according to a `mass loading' factor, $\beta$, such that
\begin{equation}
\dot{M}_{\rm eject} = \beta(V_{\rm c})\,\psi = \left(V_{\rm c}/V_{\rm SN}\right)^{-\gamma_{\rm SN}}\psi\rm.
\label{eq:sn_feedback}
\end{equation}
Here $V_{\rm c}$ is the circular velocity of the disc; $\psi$ is the star formation rate; and $V_{\rm SN}$ and $\gamma_{\rm SN}$ are adjustable parameters.  We assume $V_{\rm SN}=320$~km~s$^{-1}$ \citep{Lacey16} and $\gamma_{\rm SN}=3.4$ \citep{Baugh:2019}. The ejected gas accumulates in a reservoir of mass, $M_{\rm res}$, and then falls back within the virial radius at a rate inversely proportional to the dynamical timescale of the halo.

\subsection{Radiative Transfer}
We use the spectrophotometric radiative transfer code \grasil\ \citep{Silva98} to compute model galaxy SEDs.  Using the star formation and metal enrichment histories, gas masses and galaxy structural parameters predicted by \galform, and assuming a composition and geometry for interstellar dust, \grasil\ computes the SEDs of the model galaxies, accounting for dust extinction (absorption and scattering) of stellar radiation and its subsequent re-emission.  Here we briefly describe the \grasil\ model (for further details see \citealt{Silva98} and \citealt{Granato00}).

\grasil\ assumes that stars exist in a disc + bulge system, as is the case in \galform.  The disc has a radial and vertical exponential profile with scale-lengths, $h_{R}$ and $h_{z}$, and the bulge is described by an analytic King model profile, $\rho\propto(r^{2}+r_{\rm c}^2)^{-3/2}$ out to a truncation radius, $r_{\rm t}$. The half-mass radii, $r_{\rm disc}$ and $r_{\rm bulge}$, are predicted by \galform\ \citep[see][for more details]{Cole00}.  By definition, given the assumed profiles, the bulge core radius is related to the half-mass radius by $r_{\rm c}=r_{\rm bulge}/14.6$, whilst the radial disc scale-length, $h_{\rm R}$, is related to the disc half-mass disc radius by $h_{R}=r_{\rm disc}/1.68$.  Star formation histories are calculated separately for the disc and bulge by \galform.  For galaxies undergoing a starburst, the burst star formation, as well as the associated gas and dust, are assumed also to be in an exponential disc but with a half-mass radius $r_{\rm burst}=\eta r_{\rm bulge}$, rather than $r_{\rm disc}$, where $\eta$ is an adjustable parameter (here $\eta=1$; see \citealt{Granato00}). The disc axial ratio, $h_{z}/h_{R}$, is a parameter of the \grasil\ model; for starburst galaxies, the axial ratio of the burst is allowed to be different (0.5) from that of discs in quiescent galaxies (0.1).

The gas and dust exist in an exponential disc, with the same radial scale-length as the disc stars but in general with a different scale-height, so $h_{z}\mathrm{(dust)}/h_{z}\mathrm{(stars)}$ is an adjustable parameter.  The gas and dust are assumed to exist in two components: (i) giant molecular clouds in which stars form, escaping on some time scale, $t_{\rm esc}$; and (ii) a diffuse cirrus ISM.  The total gas mass, $M_{\rm cold}$, and gas-phase metallicity, $Z_{\rm cold}$, are calculated by \galform.  The fraction of gas in molecular clouds is determined by the parameter $f_{\rm cloud}$.  The cloud mass, $m_{\rm cloud}$, and radius, $r_{\rm cloud}$, are also parameters, though the results of the model depend only on the ratio, $m_{\rm cloud}/r_{\rm cloud}^2$, which determines (together with the gas metallicity) the optical depth of the clouds.  

The dust is assumed to consist of a mixture of graphite and silicate grains and polycyclic aromatic hydrocarbons (PAHs), each with a distribution of grain sizes.  The grain mix and size distribution were determined by Silva et al. so that the extinction and emissivity properties of the local ISM are reproduced using the optical properties of the dust grains tabulated by \cite{DraineLee84}.  At long wavelengths ($\lambda>30$~$\muup$m) this results in a dust opacity that approximates $\kappa_{\rm d}\propto\lambda^{-2}$.  However, in galaxies undergoing a starburst this is modified (for $\lambda>100$~$\muup$m) such that $\kappa_{\rm d}\propto\lambda^{-\beta_{\rm b}}$, where $\beta_{\rm b}$ is treated as an adjustable parameter.  Laboratory measurements suggest that values in the range $\beta_{\rm b}=1.5-2$ are acceptable (Agladze et al. \citeyear{Agladze96}) and more recent experiments that hotter dust favour lower values of $\beta_{\rm b}$ \citep{Boudet05}. Note that in our model burst galaxies have higher dust temperatures on average that their quiescently star-forming counterparts \citep{Cowley16SEDs}.  Here a value of $\beta_{\rm b}=1.5$ is adopted \citep{Lacey16}. The total dust mass in a galaxy is proportional to the cold gas mass and metallicity, both of which are predicted by \galform. 

We use the stellar population synthesis models of \cite{Maraston05}, adopting a \cite{Kennicutt83} IMF for stars that form quiescently and a top-heavy IMF for stars made in bursts. For calculating broadband photometry we convolve the predicted galaxy SED with the relevant filter transmission, and assume the prescription of \cite{Meiksin05} for attenuation due to the intergalactic medium. Note that we do not change any adjustable \grasil\ parameters from the values we have used in previous works (see e.g. values in Table~2 of Cowley et al. \citeyear{Cowley:2018-JWST}).  

Due to the computational expense of the radiative transfer calculation, we select a sub-sample of galaxies from \galform's original output on which to run \grasil. Similarly to \cite{Cowley:2018-JWST}, where the same model was used to make predictions for forthcoming deep galaxy surveys with the James Webb Space Telescope, we sample galaxies according to their stellar mass; however, here we also ensure that within each stellar mass bin the specific star formation rate distribution is fairly sampled.

\subsection{Calculating predicted quantities}
We now briefly explain how the model predictions are calculated from the output galaxy SEDs and how various quantities are related to each other. An observed frequency is denoted by $\nu$, which is related to the emitted frequency, $\nu_{\rm e}$, by $\nu_{\rm e}=\nu\,(1+z)$.

Once our output SEDs have been convolved with the appropriate (redshifted) filter transmission it is possible to construct the luminosity function, $\mathrm{d}n/\mathrm{d}\ln L_{\nu}$, at each output time. This is then related to the galaxy number counts, $\mathrm{d}\eta/\mathrm{d}\ln S_{\nu}$ (we use $\eta$ here to denote the surface number density of galaxies, rather than $n$ which we use for the comoving number density), by      
\begin{equation}
\frac{\mathrm{d}\eta}{\mathrm{d}\ln S_{\nu}}=\int\,\frac{\mathrm{d}n}{\mathrm{d}\ln L_{\nu(1+z)}}\frac{\mathrm{d}V}{\mathrm{d}z}\,\mathrm{d}z ,
\label{eq:ncts_lfs}
\end{equation}
where $\mathrm{d}V/\mathrm{d}z$ is the comoving volume element per unit solid angle and flux is related to luminosity according to
\begin{equation}
S_{\nu} = (1+z)\,\frac{L_{\nu(1+z)}}{4\pi\,d_{\rm L}^{2}(z)}\rm,
\end{equation}
where $d_{\rm L}(z)$ is the luminosity distance to redshift $z$. The EBL, $I_{\nu}$, the intensity per unit frequency per unit solid angle, is then simply the flux-weighted integral of the number counts   
\begin{equation}
I_{\nu} = \int\,S_{\nu}\,\frac{\mathrm{d}\eta}{\mathrm{d}\ln S_{\nu}}\,\mathrm{d}\ln S_{\nu}\rm.
\label{eq:int_ncts}
\end{equation}
The EBL can also be calculated directly from the CSED, $\varepsilon_{\nu}$, which describes the luminosity density per unit frequency at a given epoch, and is the luminosity-weighted integral of the luminosity function\footnote{It can also be thought of as the volume-weighted sum of our output \grasil\ SEDs, where the volume weights are obtained from our sampling strategy.},
\begin{equation}
\varepsilon_{\nu}=\int\,L_{\nu}\,\frac{\mathrm{d}n}{\mathrm{d}\ln L_{\nu}}\,\mathrm{d}\ln L_{\nu}\rm.
\label{eq:csed_lnu}
\end{equation} 
To obtain the EBL this is then integrated according to
\begin{equation}
I_{\nu} = \int\,(1+z)\,\frac{\varepsilon_{\nu(1+z)}}{4\pi\,d_{\rm L}^{2}(z)} \frac{\mathrm{d}V}{\mathrm{d}z}\,\mathrm{d}z\rm.
\label{eq:int_csed}
\end{equation}
The total EBL intensity (or brightness) per unit solid angle at $z=0$, $I$, is then obtained by integrating over frequency: 
\begin{equation}
I = \int \nu\,I_{\nu}\,\mathrm{d}\ln\nu\rm.
\label{eq:energy}
\end{equation}
This is often divided into the optical/near-IR intensity, $I_{\rm COB}$, and the far-IR intensity, $I_{\rm CIB}$, where the integral in Eqn.~\ref{eq:energy} is performed between $\lambda_{\rm obs}=(0.1,8)$ and $(8,1000)$~$\muup$m respectively.

\subsection{Variant models with a universal IMF}

\begin{figure}
\centering
\includegraphics[trim={0.4cm 0.5cm 0.1cm  0.5cm},clip,width=1.075\columnwidth]{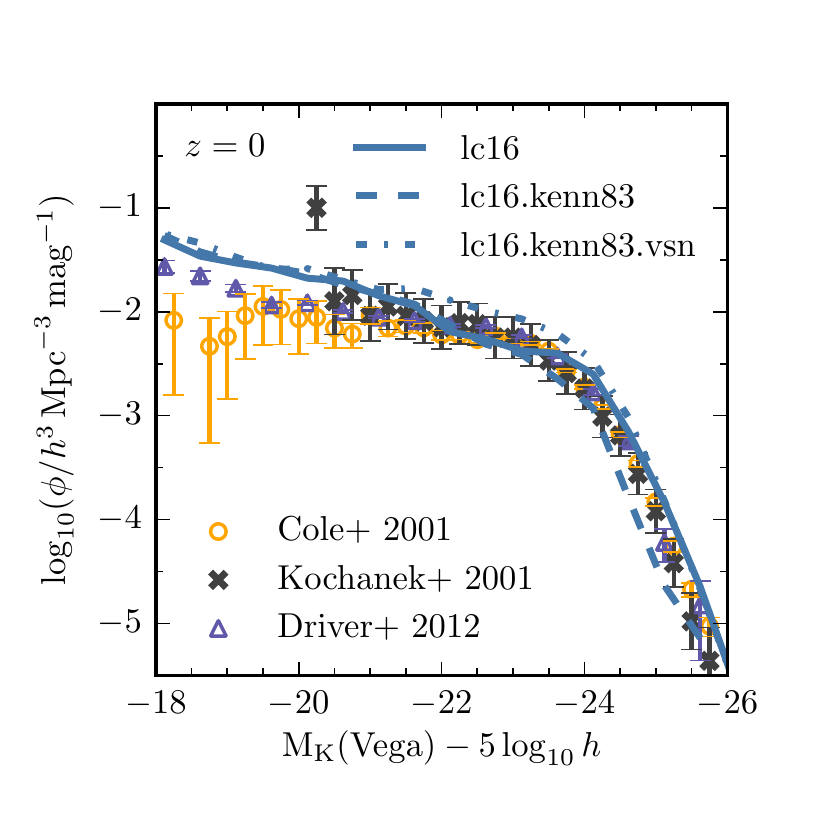}
\caption{
The $K$-band luminosity function at $z=0$. Model predictions are from the fiducial model (lc16, solid line), a variant with a universal Kennicutt (\citeyear{Kennicutt83}) IMF (lc16.kenn83, dashed line) and a variant with a universal IMF and a lower supernovae feedback mass-loading normalisation (lc16.kenn83.vsn, dash-dotted line). Luminosity function data are from Cole et al. (\citeyear{Cole:2001}), Kochanek et al. (\citeyear{Kochanek:2001}) and Driver et al. (\citeyear{Driver:2012}). 
}
\label{fig:kbandz0}
\end{figure}

We also investigate two variant {\tt GALFORM} models to illustrate the effect of relaxing the  assumption of a top-heavy IMF in bursts which is a key component of the fiducial model. Note that we do not consider either of these variants to be viable models as they fail to reproduce the calibration data to the same level as the fiducial model, as  discussed below. 

In the first variant model we turn off the top-heavy IMF option in the  fiducial model such that all stars form with a universal \cite{Kennicutt83} IMF, but leave all other parameters unchanged (this variant is labelled lc16.kenn83). The predicted $K$-band luminosity function for this model is shown in Fig.~\ref{fig:kbandz0}. This variant matches the fiducial model faintwards of $L_*$ but under-predicts the number of bright galaxies by up to a factor of two. This is because in the fiducial model only a small fraction ($\lesssim5$~per~cent) of the $z=0$ stellar mass density was formed in dynamically-triggered bursts with a top-heavy IMF \citep[e.g.][]{Gonzalez11}. 

As well as under-predicting the bright-end of the present day luminosity function, we will see later that this model dramatically fails to reproduce the mid- to far-IR EBL, which relates also to the generally poor agreement with the cosmic star formation history (see Section 3.4 and Figs 7 \& 8). To mitigate these shortcomings, we also consider another variant model in which, as well as assuming a universal IMF, we also reduce the value of the $V_{\rm SN}$ parameter, which controls the normalisation of the mass-loading factor for supernova feedback (see Eqn.~\ref{eq:sn_feedback}), from $320$~km~s$^{-1}$ to $290$~km~s$^{-1}$, resulting in the model labelled lc16.kenn83.vsn.  The predicted $K$-band luminosity function for this variant is also shown in Fig.~\ref{fig:kbandz0}; the match to the observed bright end is much better in this case. However, this variant model over-predicts the abundance of galaxies around $L_*$.

\section{Results}
\label{sec:Results}
Here we present the predicted UV-to-mm extragalactic background light spectrum, and show from which redshifts it originates (\S~\ref{sec:ebl}). We also present the predicted model number counts compared to the observational estimates compiled by \cite{Driver:2016}, and the predicted distribution of EBL emission redshifts (\S~\ref{sec:ncts}). We compare these redshift distributions at far-IR wavelengths to those inferred from CMB cross-correlations \citep{Schmidt:2015} and stacked \emph{Herschel} data \citep{Jauzac:2011,Bethermin12}.  In Section~\ref{sec:csed} we present the evolution of the CSED, $\varepsilon_{\nu}$, predicted by our model, compared with the observational estimates of \cite{Andrews:2017}. Finally, in Section~\ref{sec:top-heavys}, we review the consistency of the EBL, the $z=0$ $K$-band luminosity function and the cosmic star formation history in the context of one of the more controversial features of our model, namely a top-heavy IMF for starburst galaxies.

\subsection{The extragalactic background light}
\label{sec:ebl}

\begin{figure*}
\centering
\includegraphics[trim={0.2cm 0.0cm 0.cm  1.cm},clip,width=2.20\columnwidth]{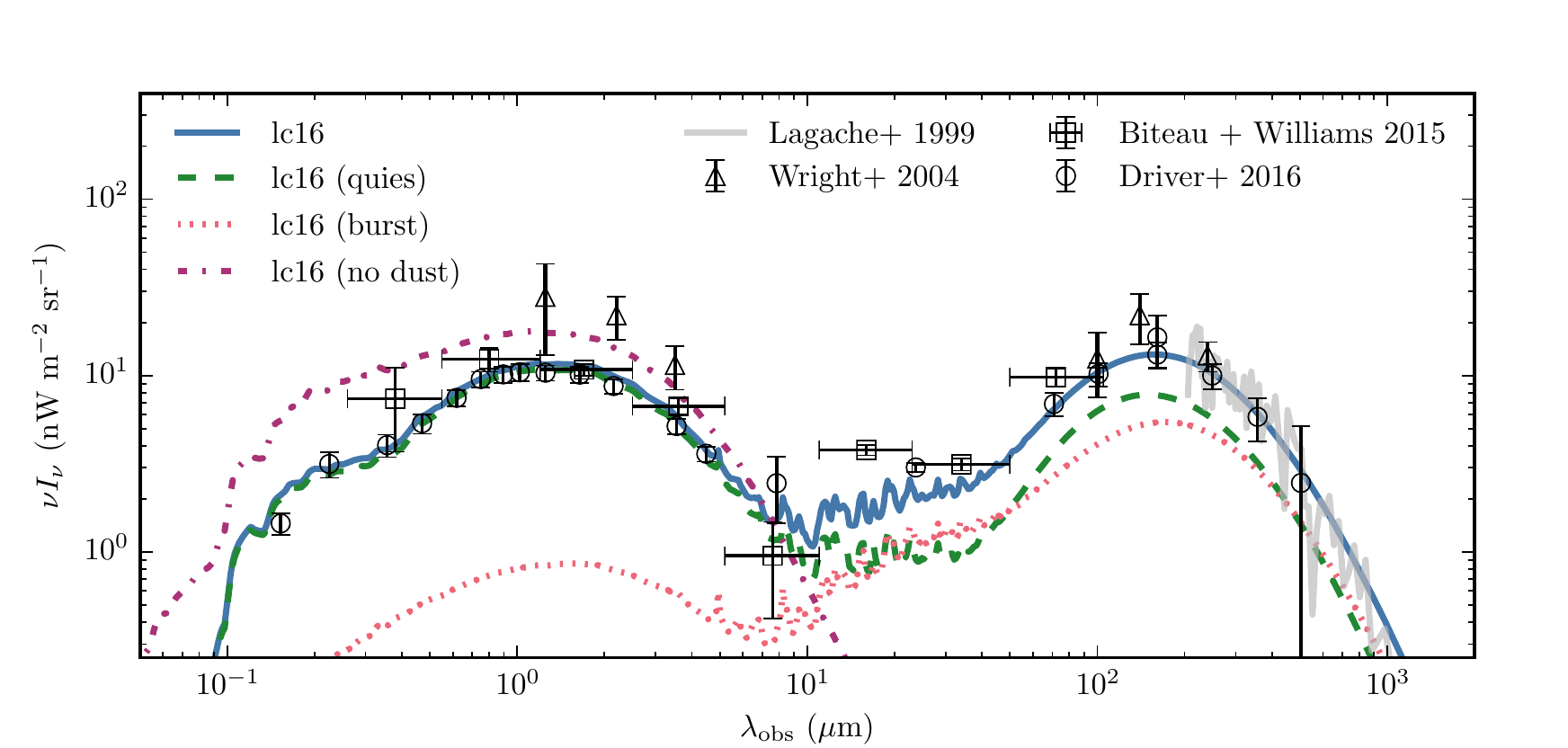}
\caption{Predicted intensity of UV-to-mm extragalactic background light (solid blue line). The contributions from quiescent and starburst galaxies are shown as dashed green and dotted red lines respectively. The EBL without the effects of dust absorption and emission is shown as the dash-dotted magenta line. Observational data are from Lagache et al. (\citeyear{Lagache:1999}, direct detection), Wright (\citeyear{Wright:2004}, direct detection), Biteau \& Williams (\citeyear{BiteauWilliams:2015}, TeV photon scattering) and Driver et al. (\citeyear{Driver:2016}, integrated number counts).}
\label{fig:ebl}
\end{figure*}

\begin{figure*}
\centering
\includegraphics[trim={0.5cm 0.0cm 0.cm  1.cm},clip,width=2.20\columnwidth]{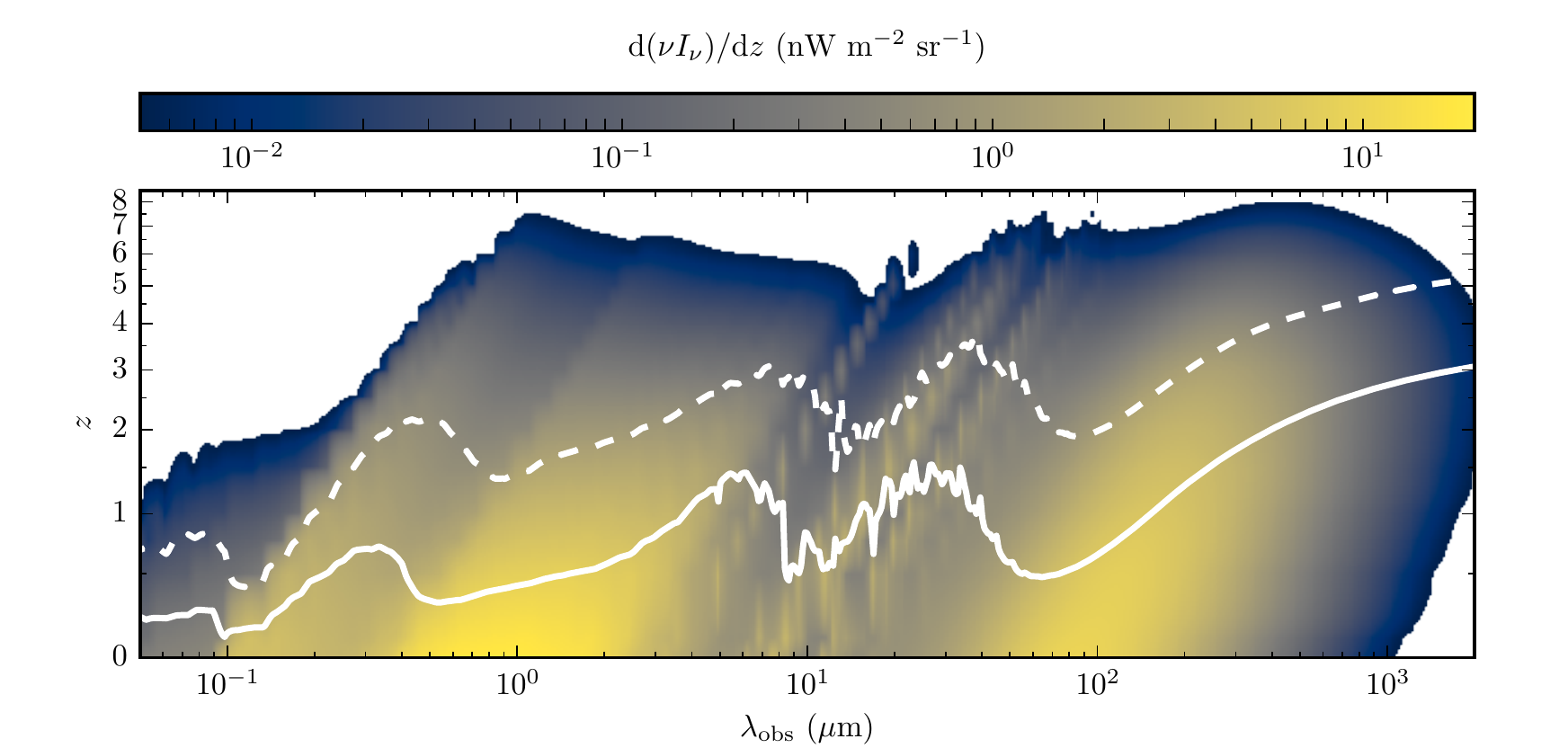}
\caption{The predicted background light per unit redshift as a function of emission redshift and observer-frame wavelength. The solid line indicates the median redshift i.e. the redshift at which half of the background light at that observed wavelength had been produced. The dashed line indicates the tenth percentile redshift. The `cividis' colormap used is described by Nu\~nez et al. (\citeyear{cividis}).}
\label{fig:ebl_dz}
\end{figure*}

The EBL predicted by our model is shown in Fig.~\ref{fig:ebl}, compared to observational data derived from a variety of methods. Different observational datasets are generally in good agreement with one another, though the discrepancy at near-IR wavelengths is evident between the direct estimates of \cite{Wright:2004}, based on data from the \emph{COBE} satellite, and the other indirect methods, as discussed in the Introduction. As described there, current data favour the estimates from galaxy number counts \citep{Driver:2016} over those from \emph{COBE} \citep{Wright:2004} at these wavelengths.

The model predictions are in excellent agreement with the observational data over the whole UV-to-mm range, with only minor discrepancies at far-UV ($\sim0.15$~$\muup$m) and mid-IR ($\sim10-30$~$\muup$m) wavelengths where the model appears to tentatively over- and under-predict the data respectively. We stress that this remarkable agreement is not a result of how the pre-existing model we are using was calibrated (see Lacey et al. \citeyear{Lacey16} for full details). 

The data used for calibration include far-IR number counts at \emph{Herschel}-SPIRE and SCUBA wavelengths ($250$, $350$, $500$ and $850$~$\muup$m), which tend to be dominated by galaxies with fluxes brighter than those that dominate the background light. This is because the galaxy counts are often determined from confusion-limited imaging at these wavelengths, which makes it difficult to resolve the fainter sources responsible for the bulk of the EBL. Additionally, the model uses the evolution of the rest-frame $K$-band luminosity function up to $z=3$ as a constraint. However, these luminosity functions span different observer-frame wavelengths at each epoch so it is not clear to what extent they constrain the EBL. 

Furthermore, the predicted EBL covers many wavelengths that were not used at all in the model calibration, as the original calibration did not include a full dust grain model and radiative transfer calculation, without which it is not possible to predict mid-IR wavelengths accurately \citep{Cowley16SEDs}. Therefore, this agreement is a genuine success of the model and reflects its predictive power, based on the self-consistent treatment of the physical processes of galaxy formation combined with the radiative transfer of stellar radiation through interstellar dust.

Interestingly, our predictions indicate that emission from quiescent galaxies (i.e. those for which star formation is not dynamically triggered and that form stars according to a solar-neighbourhood IMF) dominates the EBL over the whole UV-to-mm wavelength range, apart from at $\lambda_{\rm obs}\sim30$~$\muup$m (where there is a significant contribution from redshifted PAH emission originating in starburst galaxies) and for $\lambda_{\rm obs}\gtrsim350$~$\muup$m, where the contribution from both populations is approximately equal. It should also be noted that quiescent galaxies account for almost all of the EBL for $\lambda_{\rm obs}\lesssim8$~$\muup$m, whereas bursts make a more significant contribution at longer wavelengths. This is unsurprising, as the top-heavy IMF implemented for these galaxies during their dynamically-triggered star formation bursts is very efficient at boosting the emission from interstellar dust at these longer wavelengths \citep[][]{Baugh05}.                     

Integrating the predicted background light, we find $I_{\mathrm{COB}}=25.9$~nW~m$^{-2}$~sr$^{-1}$ ($52$~per~cent of the total), $I_{\mathrm{CIB}}=24.4$~nW~m$^{-2}$~sr$^{-1}$ ($48$~per~cent) using $8$~$\muup$m as the division between the two regimes (see Eqn.~\ref{eq:energy}). This is a very similar distribution of intensity (or brightness) to that found by observational studies \citep[e.g.][]{HauserDwek01, Dole06}, which follows from the agreement of our predictions with the observed EBL spectrum. It indicates that approximately half of the energy emitted by stars over the history of the Universe has been re-radiated by interstellar dust at longer wavelengths, and highlights the importance of understanding dust-obscured star formation for understanding the cosmic star formation history.

\cite{Andrews:2018} also briefly compared EBL predictions of the \cite{Lacey16} and \cite{vgp14} \galform\ models to observational data and the predictions of their own model. However, Andrews et al. used a slightly different version of the \cite{Lacey16} model than in this work, as the one we use here has been re-calibrated for implementation in merger trees from the P--Millennium dark matter simulation. Additionally, the \galform\ photometry in Andrews et al. was computed using the simplified dust model described in \cite{Lacey16} and not the full radiative transfer calculation with \grasil. Most of the differences between the Lacey et al. \galform\ EBL predictions at optical/near-IR wavelengths\footnote{The \galform\ line segments at $\lambda_{\rm obs}=300-400$~$\muup$m in Fig.~9 of Andrews et al. are erroneous and do not relate to the \galform\ model (Andrews et al., private communication).} presented here and those in Andrews et al. relate to the recalibration of the model, rather than the use of \grasil. We discuss this in Appendix~\ref{app:grasil_recal}.     

In Fig.~\ref{fig:ebl_dz} we show the contribution to the $z=0$ EBL from different emission redshifts. Also shown is the median emission redshift of the EBL as a function of observed wavelength, i.e. the redshift at which 50 percent of the EBL had been produced at that observed wavelength (solid line), and the redshift at which ten percent of the EBL had been produced (dashed line). From this we can see that most of the EBL is produced at $z\lesssim1$, except at $\lambda_{\rm obs}\gtrsim100$~$\muup$m, where it comes from increasingly higher redshifts as a function of increasing observed wavelength. This is a result of the negative $k$-correction \citep[e.g.][]{Blain02,Casey14} that this portion of a galaxy SED experiences. It should also be noted that the median redshift is not generally a monotonic function of wavelength, but that there are various maxima that can be related to features in the redshifted spectral energy distributions of galaxies. For example, the peak at $\lambda_{\rm obs}\sim0.3$~$\muup$m falls between the Lyman and $4000$~{\AA} breaks, the peak at $\lambda_{\rm obs}\sim5$~$\muup$m is caused by emission from old stars and the peak (and smaller features within) around $\lambda_{\rm obs}\sim30$~$\muup$m can be attributed to PAH emission.      

\subsection{Galaxy number counts and the emission redshift distribution of the EBL}
\label{sec:ncts}

\begin{figure*}
\centering
\includegraphics[trim={0.cm 0.0cm 0.cm  0.cm},clip,width=2.20\columnwidth]{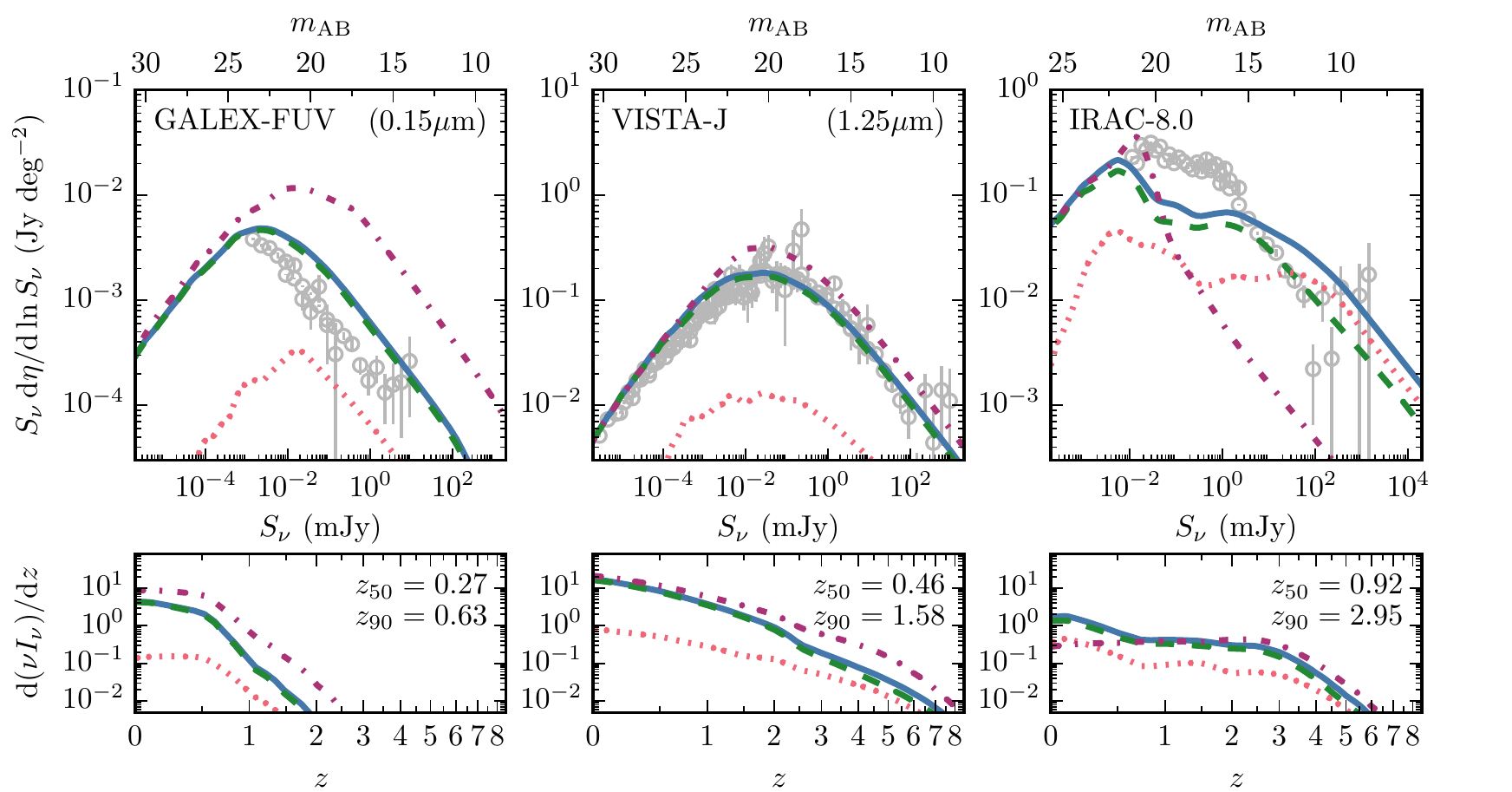}
\includegraphics[trim={0.cm 0.0cm 0.cm  0.cm},clip,width=2.20\columnwidth]{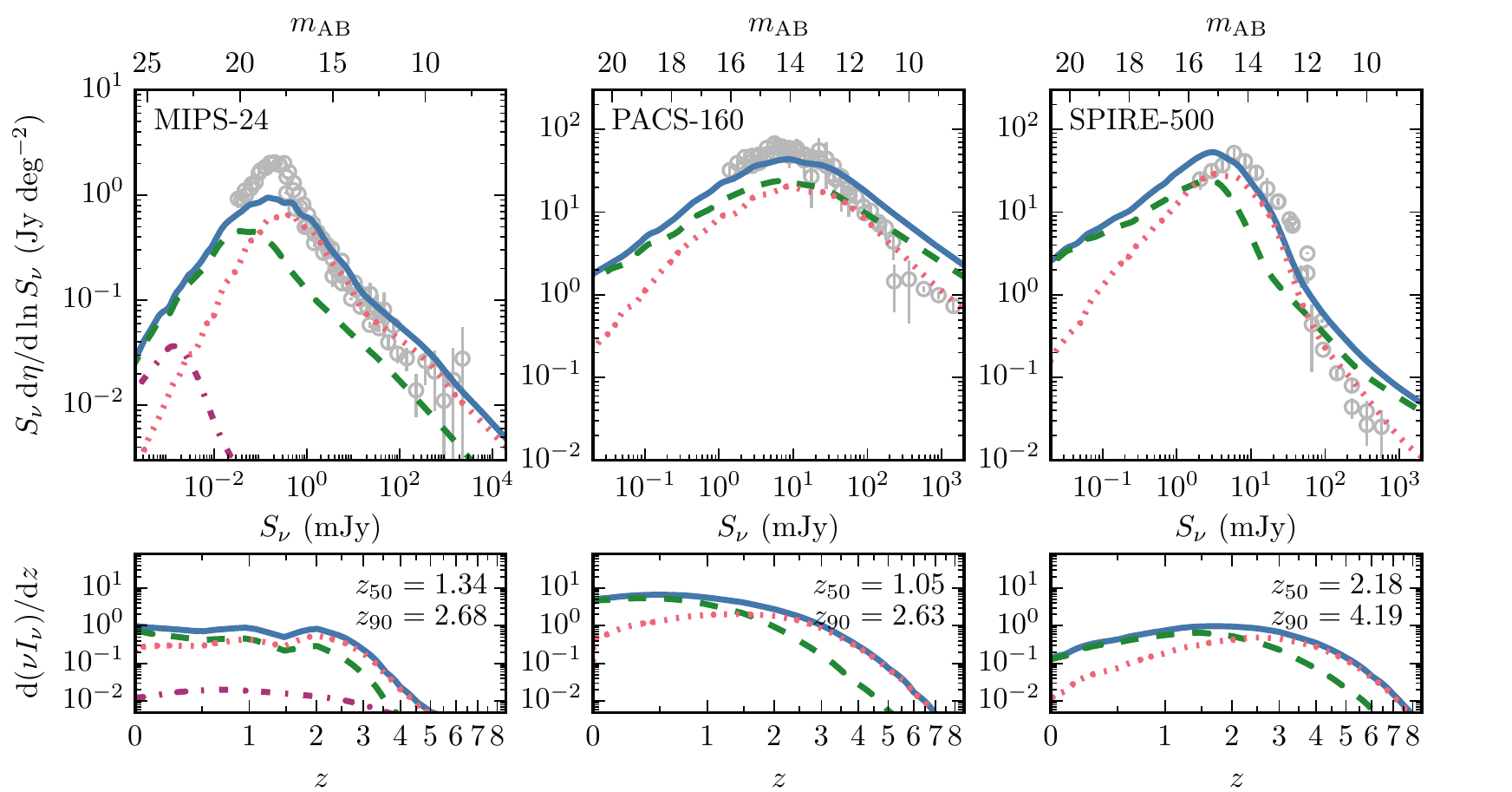}
\caption{Flux-weighted galaxy number counts (main panels) and distribution of emission redshifts for the background light (in units of nW m$^{-2}$ sr$^{-1}$). The bands are indicated in each panel. Lines have the same meaning as in Fig.~\ref{fig:ebl}. Observational data (grey points) are from the compilation of Driver et al. (\citeyear{Driver:2016}). $z_{50}$ and $z_{90}$ correspond to the median and $90^{\mathrm{th}}$ percentile redshifts of the distributions.}
\label{fig:ncts_dndz}
\end{figure*}

\begin{figure*}
\centering
\includegraphics[trim={0.cm 0.3cm 0.cm  0.9cm},clip,width=2.20\columnwidth]{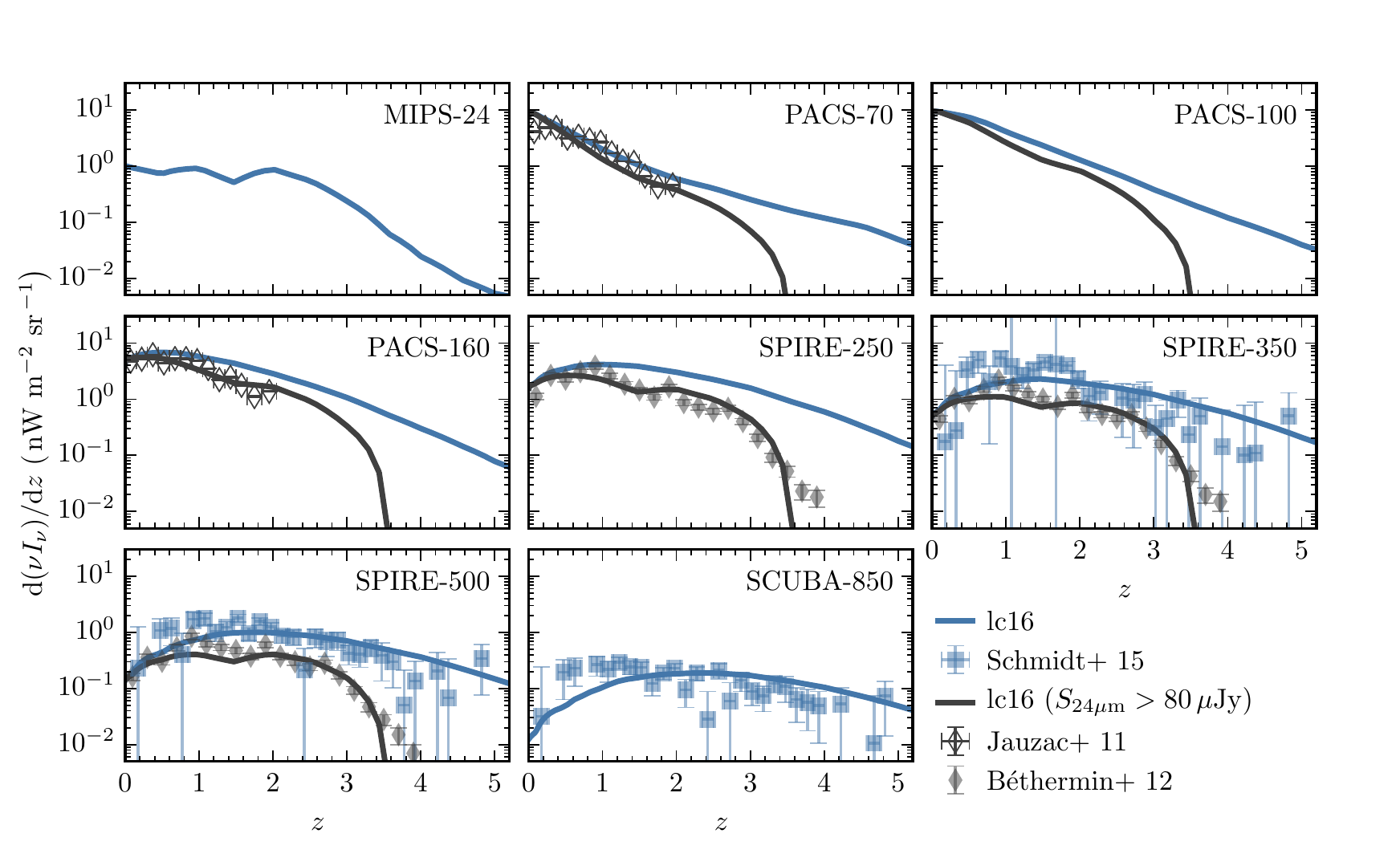}
\caption{The predicted emission redshift distribution of the background light at far-IR wavelengths for the band indicated in each panel. Our model predictions for all galaxies are the blue solid lines while predictions for galaxies with $S_{24\muup\mathrm{m}}>80$~$\muup$Jy are shown as dark grey lines. Observational data are from Jauzac et al. (\citeyear{Jauzac:2011}, open diamonds), B\'{e}thermin et al. (\citeyear{Bethermin12}, grey diamonds) and Schmidt et al. (\citeyear{Schmidt:2015}, blue squares).}
\label{fig:ebl_dndz_obs}
\end{figure*} 
As we have established the agreement between our model predictions for the EBL and current observations, it is worth investigating the agreement between the predicted and measured galaxy number counts since the background light is equal to the flux-weighted integral of the galaxy number counts, provided that all of it is emitted from galaxies. We show our predictions compared to the observational data compiled by \cite{Driver:2016} for a range of bands covering the UV-to-mm in the main panels of Fig.~\ref{fig:ncts_dndz}. The figures for other bands are shown in Appendix~\ref{app:ncts}. We have weighted the number counts by flux in these panels such that the integral under the curve with respect to the (logarithm of the) abscissa is equal to the EBL in that band (see Eqn.~\ref{eq:int_ncts}). It also allows a clear visual indication of the galaxy fluxes that contribute most to the EBL at different wavelengths. 

The agreement between our model predictions and observations is very good over the whole wavelength range. There are some small discrepancies, however. The GALEX-FUV counts appear to be over-predicted. The model also appears to under predict the peak in the counts at $\sim10^{-1}$~mJy in the IRAC-$8$~$\muup$m and MIPS-$24$~$\muup$m filters. These differences are related to the minor discrepancies seen in Fig.~\ref{fig:ebl}.

We note that the flux-weighted number counts in the SPIRE-$500$~$\muup$m filter peak at around $1$~mJy, which roughly coincides with the faintest observational data available, and that a significant proportion of the EBL comes from fainter galaxies. These faint data points are from \cite{Bethermin12}, who used a stacking analysis to derive estimates of the galaxy counts below the confusion limit of the \emph{Herschel} imaging. This highlights the point made above that calibrating the model to the bright number counts at these wavelengths does not necessarily guarantee a good agreement with the background light.

The emission redshift distribution of the background light in each band is shown in the minor panels in Fig.~\ref{fig:ncts_dndz}. We can see here that burst galaxies generally contribute more to the background light at higher redshifts, and indeed dominate the background light at mid- to far-IR wavelengths for $z\gtrsim2$. A comparison of our predictions for the redshift distribution of the background light with available observational infrared data is shown in Fig.~\ref{fig:ebl_dndz_obs}. Here we compare the observations of \cite{Schmidt:2015}, derived from cross-correlating \emph{Planck} High-Frequency Instrument maps with quasars identified in the Sloan Digital Sky Survey DR7. We find good agreement over all redshifts, though note that the errorbars on these data are significant. 

Additionally, we compare our predictions for the distribution of EBL emission redshifts to the stacked data of \cite{Jauzac:2011} and \cite{Bethermin12}. These authors stacked \emph{Herschel} images on the positions of $S_{24\muup\mathrm{m}}>80$~$\muup$Jy sources with known redshifts. We include this flux limit in our predictions and find generally good agreement with the observational data (we remind the reader that no infrared data at wavelengths shorter that $250$~$\muup$m was used in the calibration of our fiducial model). Our flux-limited predictions are slightly bi-modal. This is caused by PAH emission being redshifted through the $24$~$\muup$m filter. In this case, the relatively broad peak in the predicted $24$~$\muup$m distribution at $z\sim2$ is due to the redshifted $7.7$ and $8.6$~$\muup$m PAH features originating from starburst galaxies (see also the relevant panel in Fig.~\ref{fig:ncts_dndz}).  B\'ethermin et al. find some features in their observed distribution that are too `sharp' in redshift to be caused by PAH emission, as the width of the MIPS~$24$~$\muup$m filter causes this to appear over a broad range of redshifts. Instead, as some of these features coincide with known large-scale structures in the COSMOS field (e.g. at $z=0.3$ and $1.9$), they attribute them to cosmic structure, since the COSMOS field is small enough that the sampling variance due to cosmic structure is significant. We do not make specific predictions for the sampling variance here but reiterate that the model is able plausibly to reproduce the buildup of the infrared background light since $z\sim4$. In Appendix~\ref{app:optical_dndz} we also compare our infrared emission redshift distributions to those of \cite{Viero:2013b}, who performed a similar stacking analysis to \cite{Jauzac:2011} and \cite{Bethermin12} but instead stacked on a $K$-band selected sample and implemented a magnitude limit of $K_{\rm AB}<24$ for their procedure. Including this near-IR selection in our predictions we find a similarly good agreement with their data as that seen in Fig.~\ref{fig:ebl_dndz_obs} (see Fig.~\ref{fig:optical_dndz}).                       

\subsection{The cosmic spectral energy distribution}
\label{sec:csed}
\begin{figure*}
\centering
\includegraphics[trim={0.cm 0.3cm 0.cm  0.9cm},clip,width=2.20\columnwidth]{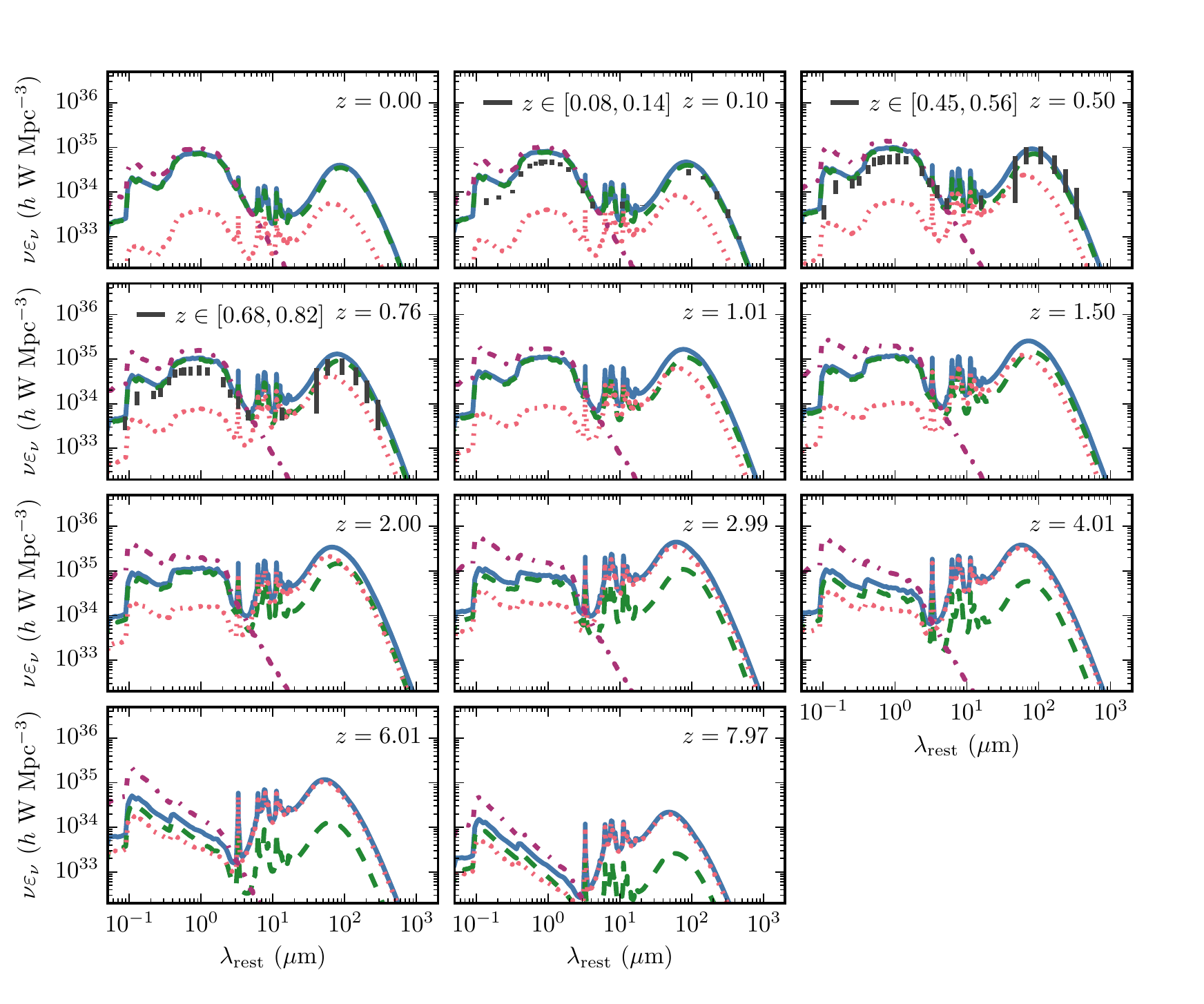}
\caption{The predicted cosmic spectral energy distribution at the redshift indicated in the panel. Model lines have the same meaning as in Fig.~\ref{fig:ebl}.  Observational data at $z<0.8$, shown as dark-grey bars, are from Andrews et al. (\citeyear{Andrews:2017}); the redshift range covered by these data is indicated in each panel.}
\label{fig:csed}
\end{figure*}
The cosmic spectral energy distribution \citep[CSED e.g.][]{Driver:2008, Andrews:2017}, $\varepsilon_{\nu}$, describes the luminosity density per unit frequency as a function of wavelength at a given epoch of the Universe's history (see Eqn.~\ref{eq:csed_lnu}). This is related to the EBL, $I_{\nu}$, which can be derived by integrating the volume-weighted (and redshifted) CSEDs over the history of the Universe (see Eqn.~\ref{eq:int_csed}).

We show the evolution of our predicted CSEDs in Fig.~\ref{fig:csed}. The optical to near-IR continuum slopes ($\lambda_{\rm rest}\sim0.4-3$~$\muup$m) evolve quite dramatically from $z=8$ to $z\sim3$, as at these wavelengths the buildup of old stars contributes to a flatter spectrum by $z\sim3$. This is independent of dust attenuation, as we can observe a similar evolution in the unattenuated CSEDs. The UV continuum slopes appear to remain fairly blue [i.e. $\beta_{\rm UV}\sim-2$ if we fit the far-UV ($0.1<\lambda_{\rm rest}<0.3$~$\muup$m) portion of the CSED with a powerlaw, $\varepsilon_{\nu}\propto\lambda^{2+\beta_{\rm UV}}$] at all redshifts, which may contribute in part to the over-prediction of the EBL at far-UV wavelengths. The far-IR emission is dominated by burst galaxies for $z\gtrsim2$, and they continue to play a prominent role in the average PAH emission until $z\sim0.5$, but never make a significant contribution at shorter wavelengths due to a greater dust attenuation in bursts.  

We compare our predictions to the observational estimates of \cite{Andrews:2017} for $z<0.8$, where our simulation snapshots coincide with their redshift bins. Andrews et al. estimated the CSED by summing fitted SEDs based on {\sc magphys} templates \citep{daCunha:2008} and photometry from the Galaxy and Mass Assembly survey (GAMA, Driver et al. \citeyear{Driver:2011}) for $0.02<z<0.2$ and the G10 region of the Cosmic Evolution Survey (COSMOS, Scoville et al. \citeyear{Scoville:2007}) for $0.2<z<1.0$. Andrews et~al. assume a standard solar neighbourhood IMF in their analysis. We do not expect that this choice will have a significant  impact on the CSED they recover, however, as observations over a wide range of wavelengths are used to constrain the SED fitting, including the far-infrared and sub-millimetre range.\footnote{ The choice of IMF will, however, have an effect on the values of the derived physical parameters, such as the star formation rate, in the models used by Andrews et~al. Our main interest is in the form of the CSED that they recover, and since this is constrained to match observations over the UV-mm wavelength range it should not be strongly affected by the choice of IMF.} Here we show their strict lower and upper bounds [columns labelled (1) and (3) respectively in their Tables 1 and 2], which are respectively the $V_{\rm max}$ corrected sum of their galaxy {\sc magphys} SEDs and the $V_{\rm max}$ corrected sum with a spline-based optical luminosity completeness correction and upper limits included. There is generally good agreement between our predictions and the Andrews et al. estimates, particularly in the far-IR. The model, however, does seem to mildly over-predict the optical emission in this redshift range, and the predicted UV continuum slopes appear to be too steep (i.e. too blue) relative to observations, which is probably connected to our over-prediction of the GALEX-FUV number counts seen in Fig.~\ref{fig:ncts_dndz}.

\subsection{The importance of a top-heavy IMF}
\label{sec:top-heavys}

\begin{figure}
\centering
\includegraphics[trim={0.cm 0.5cm 0.1cm  0.5cm},clip,width=1.075\columnwidth]{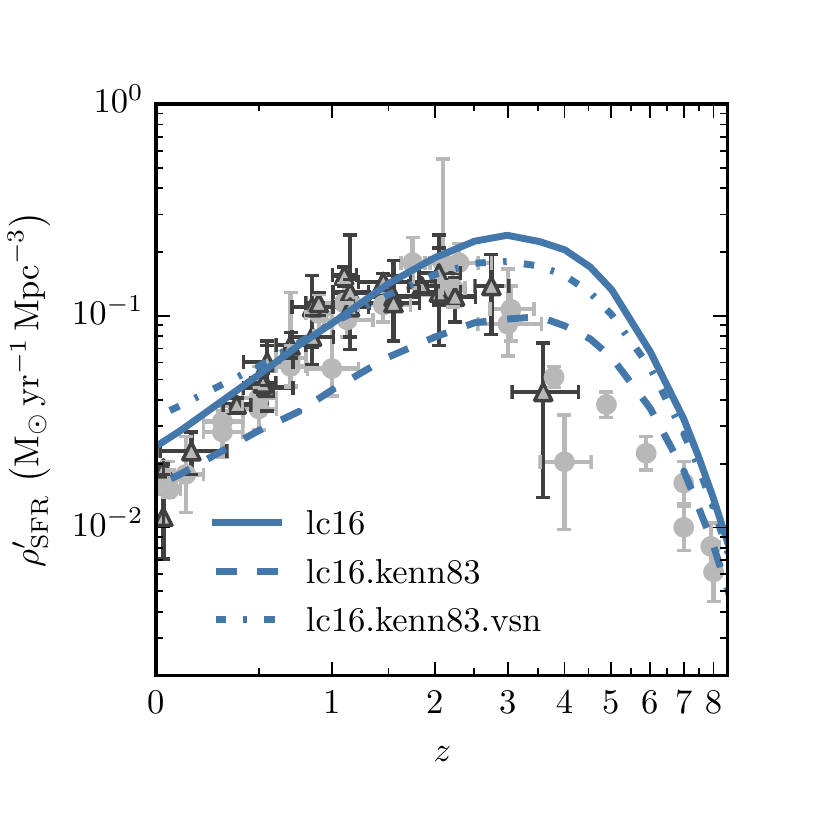} 
\caption{The apparent cosmic star formation history. Model predictions are from the fiducial model (lc16, solid line), a variant with a universal Kennicutt (\citeyear{Kennicutt83}) IMF (lc16.kenn83, dashed line) and a variant with a universal IMF and a lower supernovae feedback mass-loading normalisation (lc16.kenn83.vsn, dot-dashed line). Cosmic star formation history data are as compiled by Madau \& Dickinson (\citeyear{MadauDickinson:2014}), data from UV and far-IR surveys are shown as light grey circles and light grey triangles with a darker outline respectively. The model predictions for $\rho_{\rm SFR}^{\prime}$ are calculated using Eqn.~\ref{eq:Madau_SFR}.}
\label{fig:csfrd}
\end{figure}

\begin{figure*}
\centering
\includegraphics[trim={0.cm 0.0cm 0.cm  0.9cm},clip,width=2.2\columnwidth]{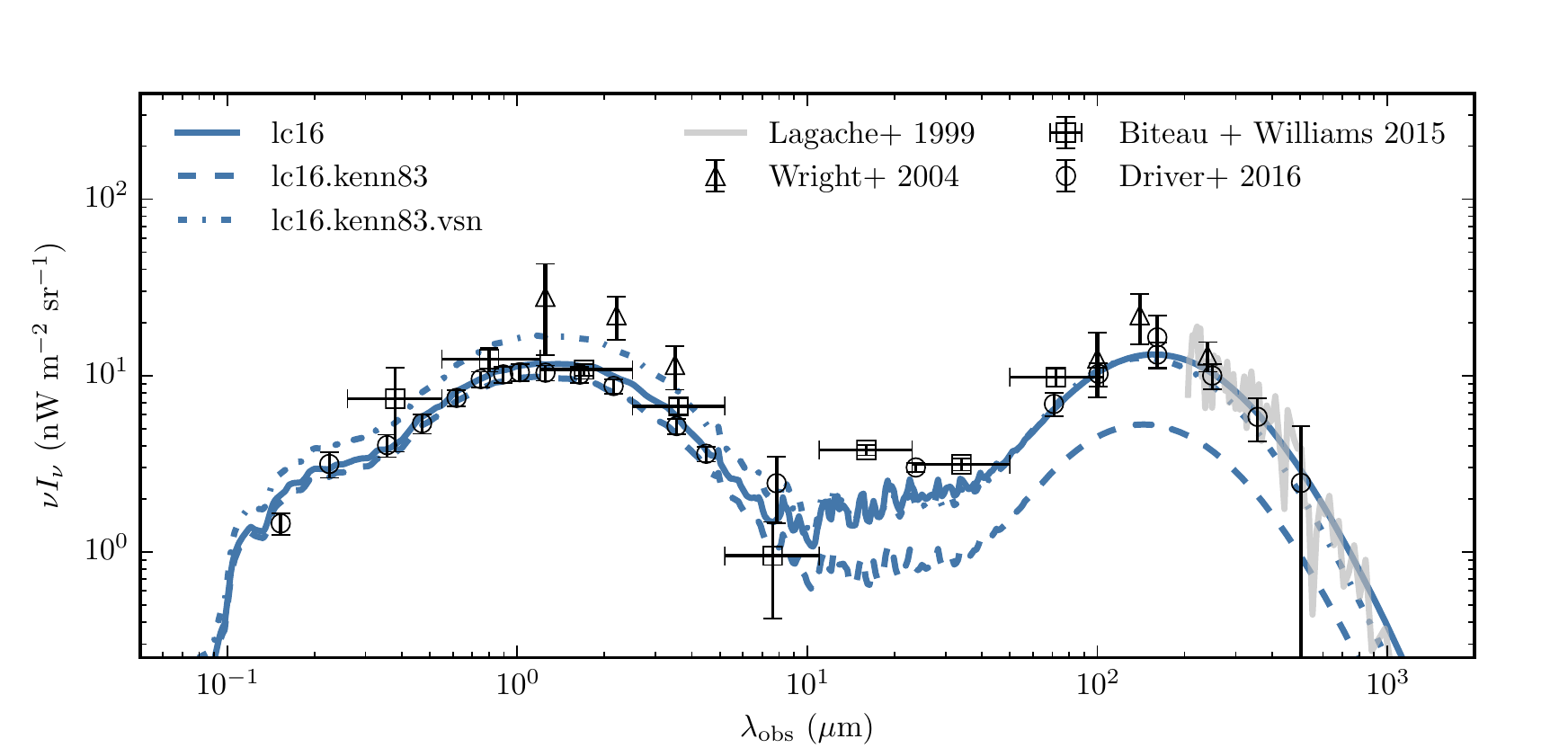}
\caption{The extragalactic background light. Model predictions are from the fiducial model (lc16, solid line), a variant with a universal Kennicutt (\citeyear{Kennicutt83}) IMF (lc16.kenn83, dashed line) and a variant with a universal IMF and a lower supernovae feedback mass-loading normalisation (lc16.kenn83.vsn, dash-dotted line). EBL data are as in Fig.~\ref{fig:ebl}.}
\label{fig:imf}
\end{figure*}

\begin{figure}
\centering
\includegraphics[trim={0.cm 0.5cm 0.1cm  0.5cm},clip,width=1.075\columnwidth]{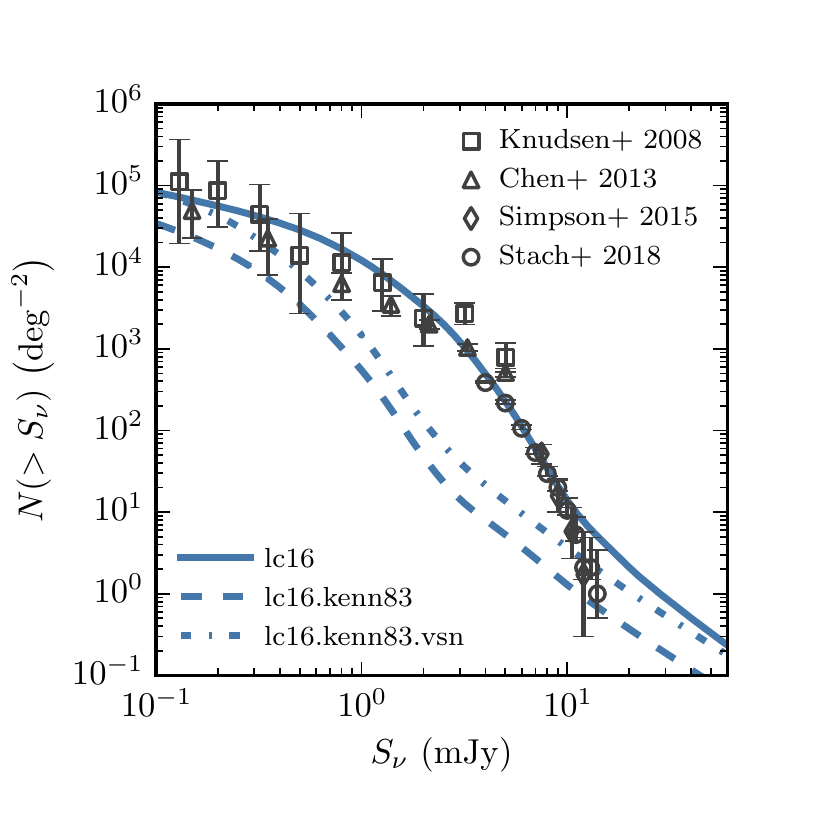}
\includegraphics[trim={0.cm 0.5cm 0.1cm  0.5cm},clip,width=1.075\columnwidth]{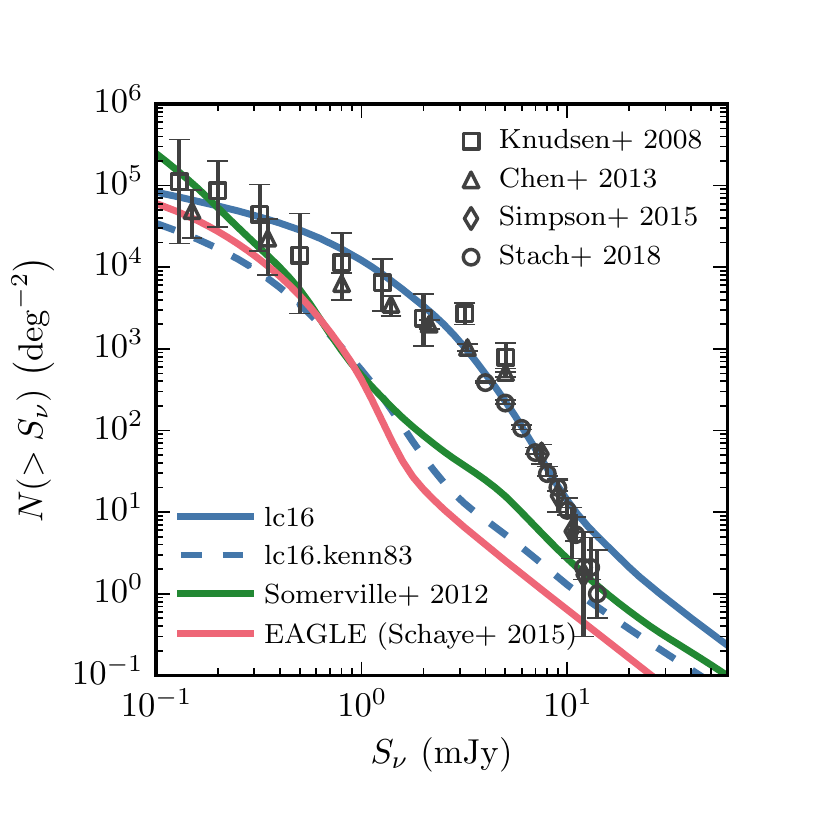}
\caption{Predicted galaxy number counts at $850$~$\muup$m. Model lines in the top panel are as in Fig.~\ref{fig:imf}. Lensed single-dish observational data are from Knudsen et al. (\citeyear{Knudsen08}, squares) and Chen et al. (\citeyear{Chen13}, triangles) and are shown only for $S_{850~\muup\mathrm{m}}\leq5$~mJy. Interferometric data are from Simpson et al. (\citeyear{Simpson15-ncts}, diamonds) and Stach et al. (\citeyear{Stach:2018}, circles). In the bottom panel the predictions from the Somerville et al. (\citeyear{Somerville:2012}, green line) semi-analytical model and {\sc eagle} hydrodynamical simulation (Schaye et al. \citeyear{Schaye15}, red line) are also shown.}
\label{fig:smg_ncts}
\end{figure}

We now investigate the extent to which the ability to reproduce the EBL relies on the top-heavy IMF assumed for dynamically-triggered star formation in our model. In doing so we reassess the argument first put forward by \cite{Fardal:2007}, namely that the present day stellar mass density, the cosmic star formation history and the EBL are not consistent with one another if a uniform solar-neighbourhood IMF is assumed. Fardal et al. argued that an IMF that is ``paunchy'' on average, containing an excess of stars in the range $1<m<8$~M$_{\sun}$ (see their Table~1 for a precise definition), is most favoured by these observational constraints.

Here we investigate this using our model. As the form of the IMF is an assumption made in observational estimates of physical properties such as stellar mass and star formation rate, and that is precisely what we are trying to investigate here, we compare only to directly observable properties. As a proxy for local stellar mass density we use the $K$-band luminosity function at $z=0$ (see the comparison of the predictions of the fiducial model and the two variants considered here with observational estimates in Fig.~\ref{fig:kbandz0})\footnote{We stress that the $K$-band luminosity function shown in Fig.~\ref{fig:kbandz0} was assigned the most weight, amongst various observational constraints, in calibrating the fiducial model.} and for the cosmic star formation history we use the data compilation of \cite{MadauDickinson:2014} . However, in the latter case, rather than comparing the star formation rates predicted directly by our model, we compute the predicted IR ($8-1000$~$\muup$m) and attenuated far-UV luminosity densities ($\rho_{\rm IR}$ and $\rho_{\rm FUV,atten}$ respectively) and convert these into apparent star formation rates using the same conversion factors as \citeauthor{MadauDickinson:2014} (see their Eqn.~12) i.e.
\begin{equation}
\rho^{\prime}_{\rm SFR} = \kappa_{\rm FUV}\,\rho_{\rm FUV,atten} + \kappa_{\rm IR}\,\rho_{\rm IR} ,
\label{eq:Madau_SFR}
\end{equation}
where $\kappa_{\rm FUV}=1.3\times10^{-28}$~M$_{\sun}$~yr$^{-1}$~erg$^{-1}$~s~Hz and $\kappa_{\rm IR}=4.5\times10^{-44}$~M$_{\sun}$~yr$^{-1}$~erg$^{-1}$~s, which are derived for a Salpeter IMF with $0.1<m<100$~M$_{\sun}$. This is not the same as the intrinsic star formation rate density predicted by \galform, which is why we use the prime symbol, $\prime$, to denote the apparent cosmic star formation history, $\rho^{\prime}_{\rm SFR}$, in Eqn.~\ref{eq:Madau_SFR}. This is discussed in more detail in Appendix~\ref{app:csfrd}.

Our model predictions for the apparent star formation rate density are compared to observational data in  Fig.~\ref{fig:csfrd} which shows that the model can reproduce the cosmic star formation rate density reasonably well for $z\lesssim2.5$. The variant with a universal IMF and the same SN feedback as the fiducial model under-predicts the apparent star formation rate density at $z<2$. The variant with reduced SN feedback gives a similar prediction to the fiducial model. The fiducial model appears to over-predict the observational data at higher redshifts. However, in this redshift regime the data are highly uncertain, as most of the observational constraints come from far-UV luminosity functions and so are sensitive to assumptions made about dust attenuation and also typically involve large  extrapolations of observed far-UV luminosity functions to fainter magnitudes than actually probed by the data. According to our model, the apparent star formation rate density at these redshifts is dominated by dust-obscured star formation (see Fig.~\ref{fig:csfrd_app}). Complementary far-IR observations are extremely challenging at these redshifts, as the coarse angular resolution of single-dish telescopes used for imaging surveys at these wavelengths means that it is only possible to resolve the most highly star-forming objects. It is therefore possible that a significant amount of infrared luminosity density is currently unaccounted for at $z\gtrsim3$ \citep[e.g.][]{RowanRobinson:2016}, though this conclusion  remains controversial \citep[e.g.][]{Bouwens:2015,Finkelstein15,Bouwens:2016,Dunlop:2017}. 

Fig.~\ref{fig:imf} compares the predictions of the fiducial model and the two variants with a universal IMF to the observational estimates of the EBL.  The variant with a universal IMF and the same SN feedback as the fiducial model (lc16.kenn83) does not match the EBL observations as well as the fiducial one. Beyond $10$~$\muup$m, this variant model underpredicts the EBL by a factor of three. The variant with reduced SN feedback (lc16.kenn83.vsn) fares better at these long wavelengths. However, this model overpredicts the EBL in the near-infrared, optical and ultra-violet. Here, we remind the reader that it is most appropriate to compare our predictions with the galaxy count-based estimates of \cite{Driver:2016}. The two variants considered here therefore predict the wrong shape for the EBL.

Whilst a more detailed parameter space exploration might yield better fitting universal IMF variant models (here we have only considered varying a single parameter), it appears unlikely that they will be able to achieve as good a level of agreement as our fiducial model. In any case, the difficulties of universal IMF models in reproducing the observed abundance of bright sub-mm galaxies (whilst simultaneously reproducing other observational data) will almost certainly remain. Reproducing the abundance of bright sub-mm galaxies \citep[e.g.][]{Smail97,Hughes98} was the primary reason a top-heavy IMF was originally introduced into the \galform\ model by \cite{Baugh05}. A top-heavy IMF is extremely efficient at boosting a galaxy's sub-mm flux as: (i) more massive stars are formed per unit star formation so that the intrinsic UV luminosity is increased; and (ii) more metals and hence interstellar dust are produced through an increased supernova rate with which to absorb and re-radiate the increased amount of UV radiation at sub-mm wavelengths. Whilst the current model assumes a less top-heavy IMF than used by Baugh et al. we highlight that this feature is still necessary for this purpose in Fig.~\ref{fig:smg_ncts}, showing the number counts of sub-mm galaxies for the different models. The universal IMF variants dramatically fail to reproduce the abundance of bright ($\sim1-10$~mJy) sub-mm galaxies by around an order of magnitude, and it is difficult to see how further reducing the impact of feedback mechanisms in the model would solve this problem without the predictions becoming inconsistent with the $z=0$ distribution of $K$-band light. 

We add that this is not a difficulty unique to the \galform\ model, but is shared by other semi-analytical models \citep[e.g.][]{Somerville:2012} and hydrodynamical simulations \citep[e.g. {\sc eagle}:][]{Schaye15, Crain:2015}, as is shown in the bottom panel of Fig.~\ref{fig:smg_ncts}. The Somerville et al. IR luminosities were predicted using the observationally calibrated dust emission templates of \cite{Rieke:2009}. The {\sc eagle} luminosities were computed using the radiative transfer code {\sc skirt} \citep[e.g.][]{Baes:2011} as described in \cite{Camps:2018} and are publicly accessible via the {\sc eagle} database\footnote{\url{http://icc.dur.ac.uk/Eagle/database.php}} \citep{McAlpine:2016}. We recognise that there are some models in the literature that claim to be able to simultaneously reproduce the sub-mm number counts as well as other observables such as the present-day galaxy stellar mass function using a universal IMF \citep[e.g.][]{Safarzadeh:2017}, and we discuss these claims in more detail in Appendix~\ref{app:grasil_v_h11}.     

Although still seen as controversial, there is a growing body of observational evidence that supports not only a non-universal IMF, but the value of the IMF slope proposed by the fiducial model in highly star-forming galaxies (e.g. Gunawardhana et al. \citeyear{Gunawardhana:2011}; Finkelstein et al. \citeyear{Finkelstein:2011}; Romano et al. \citeyear{Romano:2017}; Schneider et al. \citeyear{Schneider:2018}; Zhang et al. \citeyear{Zhang18}), as discussed in Section~\ref{sec:SF_IMF}.                       

\section{Summary}
\label{sec:Conclusion}
We have investigated the extragalactic background light (EBL) predicted by the semi-analytical galaxy formation model \galform. The model is implemented in halo merger trees from the P--Millennium, a large ($800$~Mpc)$^{3}$ cosmological $N$-body simulation \citep{Baugh:2019} run with cosmological parameters consistent with the \emph{Planck} satellite data, and is calibrated to reproduce an unprecedentedly large set of observational data at $z\lesssim6$ \citep{Lacey16}. For computing simulated galaxy SEDs accounting for the absorption and re-emission of stellar radiation by interstellar dust we combined \galform\ with the fully self-consistent radiative transfer code \grasil\ \citep{Silva98}.

The predicted EBL is in remarkable agreement with available observations over the whole of the UV-to-mm range investigated. We show that most (c. $90$~per~cent) of the EBL is produced at $z\lesssim2$, although far-IR EBL photons tend to be produced at slightly higher redshifts. Comparing the model predictions for galaxy number counts with observations, we find that the model can generally reproduce the observed distribution of fluxes well over the whole range of wavelengths. We also find that the redshift distribution of the EBL is in good agreement with observational estimates at far-IR wavelengths, and this is also the case if $24$~$\muup$m and near-IR flux limits on stacked data are considered. We show the predicted evolution of the cosmic SED, the luminosity density as a function of wavelength at a given epoch in the Universe's history. We find that this is in good agreement with available observational data at $z\lesssim1$, although the predicted UV continuum slopes appear to be too `blue' at these redshifts, and the luminosity density in the optical ($\lambda_{\rm rest}\sim0.3-4$~$\muup$m) portion of the spectrum appears to be mildly over-predicted, perhaps as a consequence of having slightly too much star formation at higher redshifts.

Finally, we investigated the necessity of a top-heavy IMF during dynamically-triggered star formation for reproducing the EBL, simultaneously with the $K$-band luminosity function at $z=0$, the cosmic star formation history and the number counts and redshift distribution of galaxies at $850 \, \mu $m, by examining the predictions of variant models with a universal IMF and comparing these with the predictions of our fiducial model. We find that variant models with a universal solar-neighbourhood IMF struggle to reproduce these observational constraints to the same level of accuracy. In particular, the universal IMF variants do not reproduce the sub-mm  ($850$~$\muup$m) galaxy number counts as well as our fiducial model, failing by over an order of magnitude. This is a challenge shared by other physical galaxy formation models. Whilst we have only investigated a small number of variant models with a universal IMF it is difficult to see how simple parameter variations in current models can alleviate this {\it mismatch whilst simultaneously} reproducing constraints such as the $K$-band luminosity function at $z=0$. Similar conclusions were reached by \cite{Somerville:2012} who, using a different semi-analyical model of galaxy formation, were unable to reproduce the counts of galaxies at $850 \, \mu$m using a model with a universal IMF. Thus it seems that these data favour a top-heavy IMF in highly star-forming galaxies, a feature which remains controversial but for which there is mounting evidence from independent observational probes \citep[e.g.][]{Zhang18}.   

The overall excellent agreement of the predictions of our pre-existing galaxy formation model with EBL data is a remarkable success of the model.  These data encode multiple aspects of the galaxy formation process and are distinct from the data originally used to calibrate the fiducial model originally. No model parameters were adjusted for the comparisons presented in this study. This work highlights the predictive power and realism of this self-consistent multi-wavelength physical model and underlines its utility as a powerful tool for interpreting and understanding multi-wavelength observational data over a broad range of redshifts.    
         
\section*{Acknowledgements}
The authors would like to thank Alessandro Bressan, Gian-Luigi Granato and Laura Silva for use of, and discussions relating to, the \grasil\ code.  Additionally, we would like to acknowledge technical assistance from Joseph Earl, Lydia Heck, John Helly and Luiz Felippe Rodrigues and helpful discussions with Marta Silva and Lingyu Wang. This work has made use of the open-source \textsc{python} packages: \textsc{numpy} \citep{numpy}, \textsc{scipy} \citep{scipy}, \textsc{matplotlib} \citep{Hunter:2007} and \textsc{ipython} \citep{ipython}. The color-vision-deficiency-friendly colours used throughout can be found at \url{https://personal.sron.nl/~pault/}.
WIC acknowledges financial support through the European Research Council Consolidator Grant ID 681627 BUILDUP. This work was supported by the Science and Technology Facilities Council [ST/K501979/1, ST/L00075X/1]. CMB acknowledges the receipt of a Leverhulme Trust Research Fellowship. CSF acknowledges an ERC Advanced Investigator Grant, COSMIWAY [GA 267291] and the Science and Technology Facilities Council [ST/F001166/1, ST/I00162X/1]. CdPL has received funding from a Discovery Early Career Researcher Award (DE150100618) and the ARC Centre of Excellence for All Sky Astrophysics in 3 Dimensions (ASTRO 3D), through project number CE170100013. 
This work used the DiRAC@Durham facility managed by the Institute for Computational Cosmology on behalf of the STFC DiRAC HPC Facility (www.dirac.ac.uk). The equipment was funded by BEIS capital funding via STFC capital grants ST/P002293/1, ST/R002371/1 and ST/S002502/1, Durham University and STFC operations grant ST/R000832/1. DiRAC is   part of the National e-Infrastructure. The {\sc eagle} simulations were performed using the DiRAC-2 facility at Durham, managed by the ICC, and the PRACE facility Curie based in France at Tr\`es Grand Centre de Calcul, CEA, Bruy\'eres-le-Ch\^{a}tel. 

\bibliographystyle{mnras}
\bibliography{ref}

%%%%%%%%%%%%%%%%%%%%%%%%%%%%%%%%%%%%%%%%%%%%%%%%%%
%%%%%%%%%%%%%%%%% APPENDICES %%%%%%%%%%%%%%%%%%%%%
\appendix
%If you want to present additional material which would interrupt the flow of the main paper, it can be placed in an Appendix which appears after the list of references.
             
\section{Galaxy number counts and the emission redshift distribution of the EBL (II)}
\label{app:ncts}
\begin{figure*}
\centering
\includegraphics[width=5.928937in]{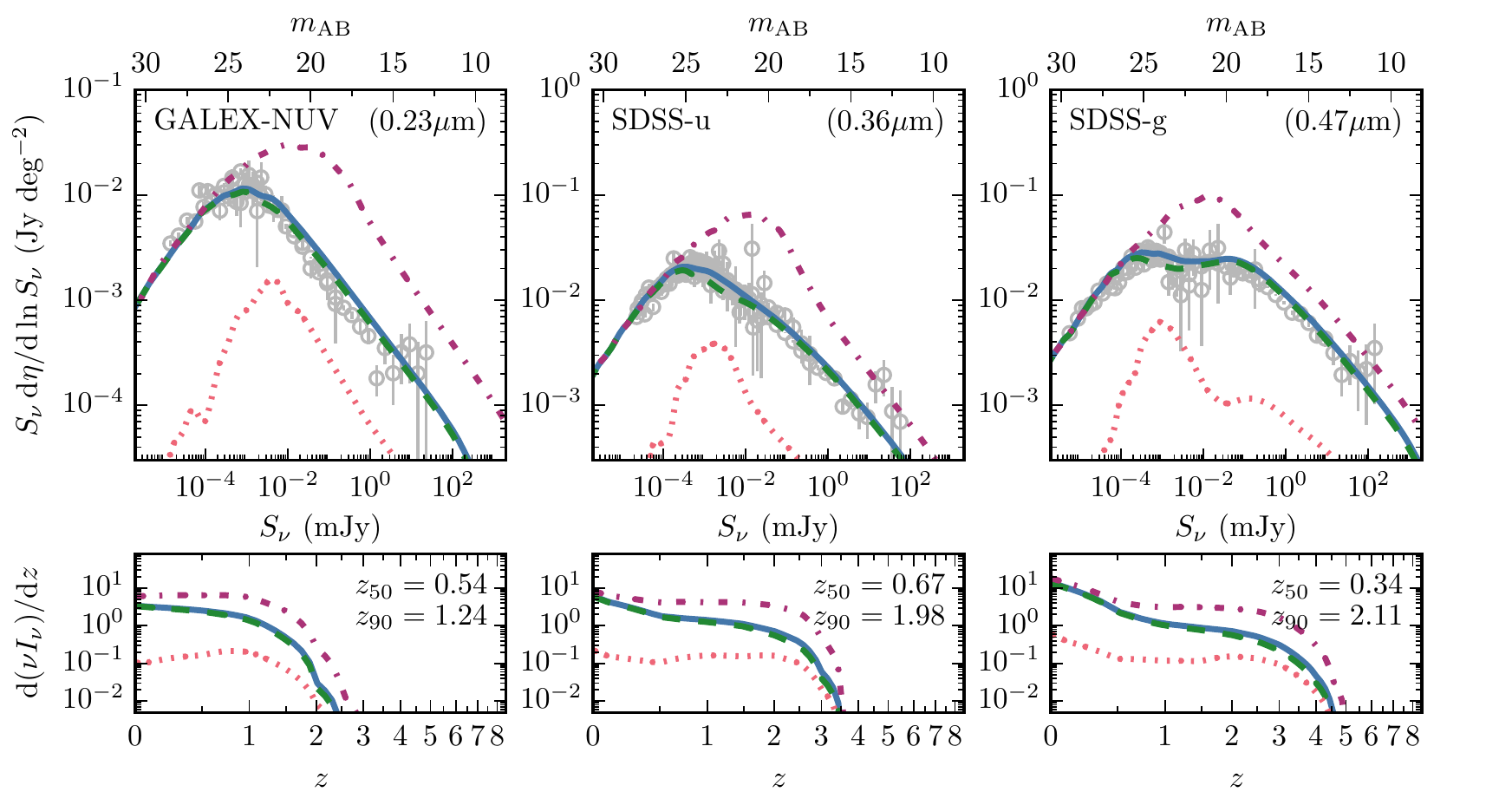}
\includegraphics[trim = 0 0 0 0, clip = true,width=5.928937in]{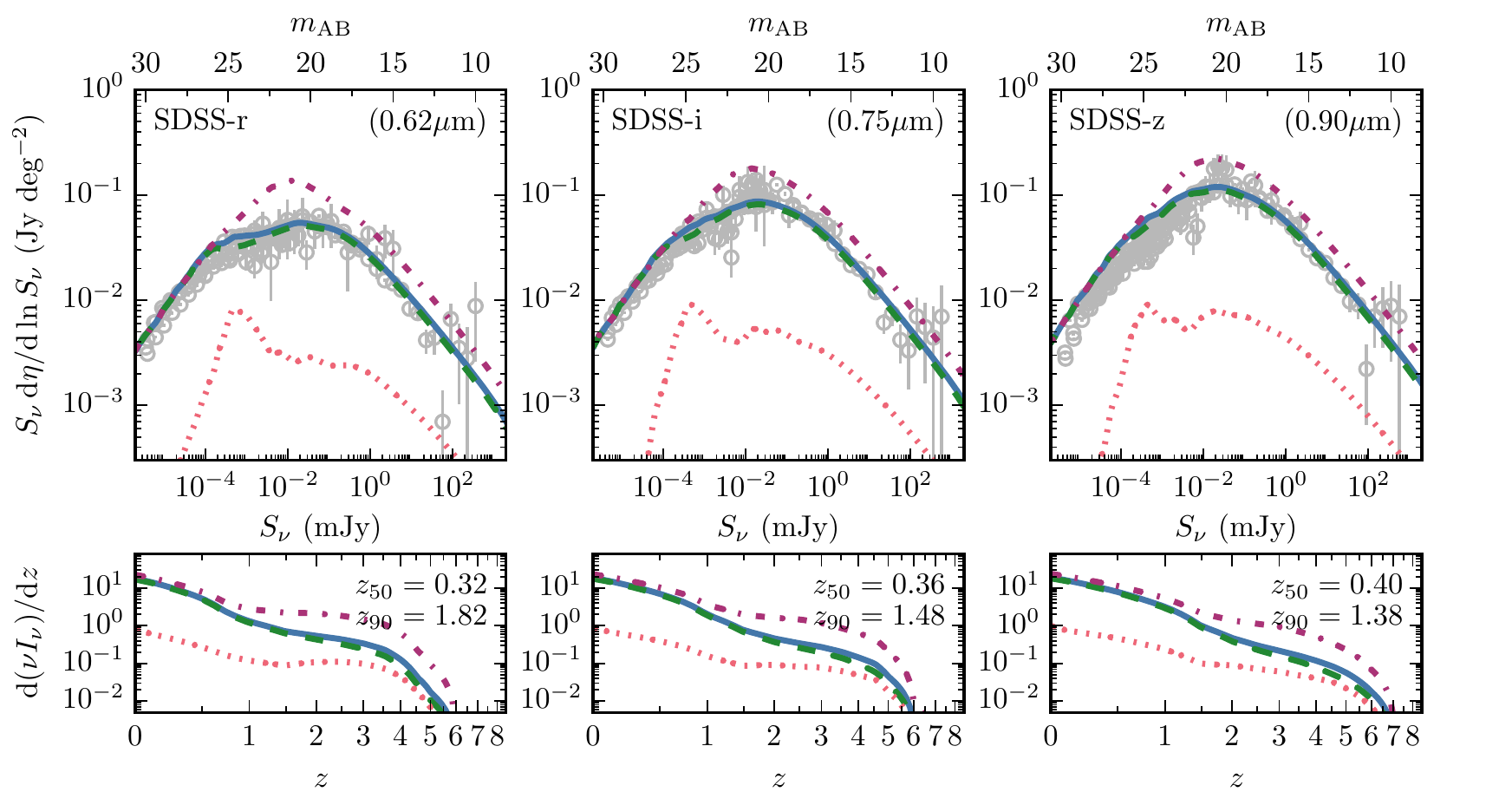}
\includegraphics[trim = 0 0 0 0, clip = true,width=5.928937in]{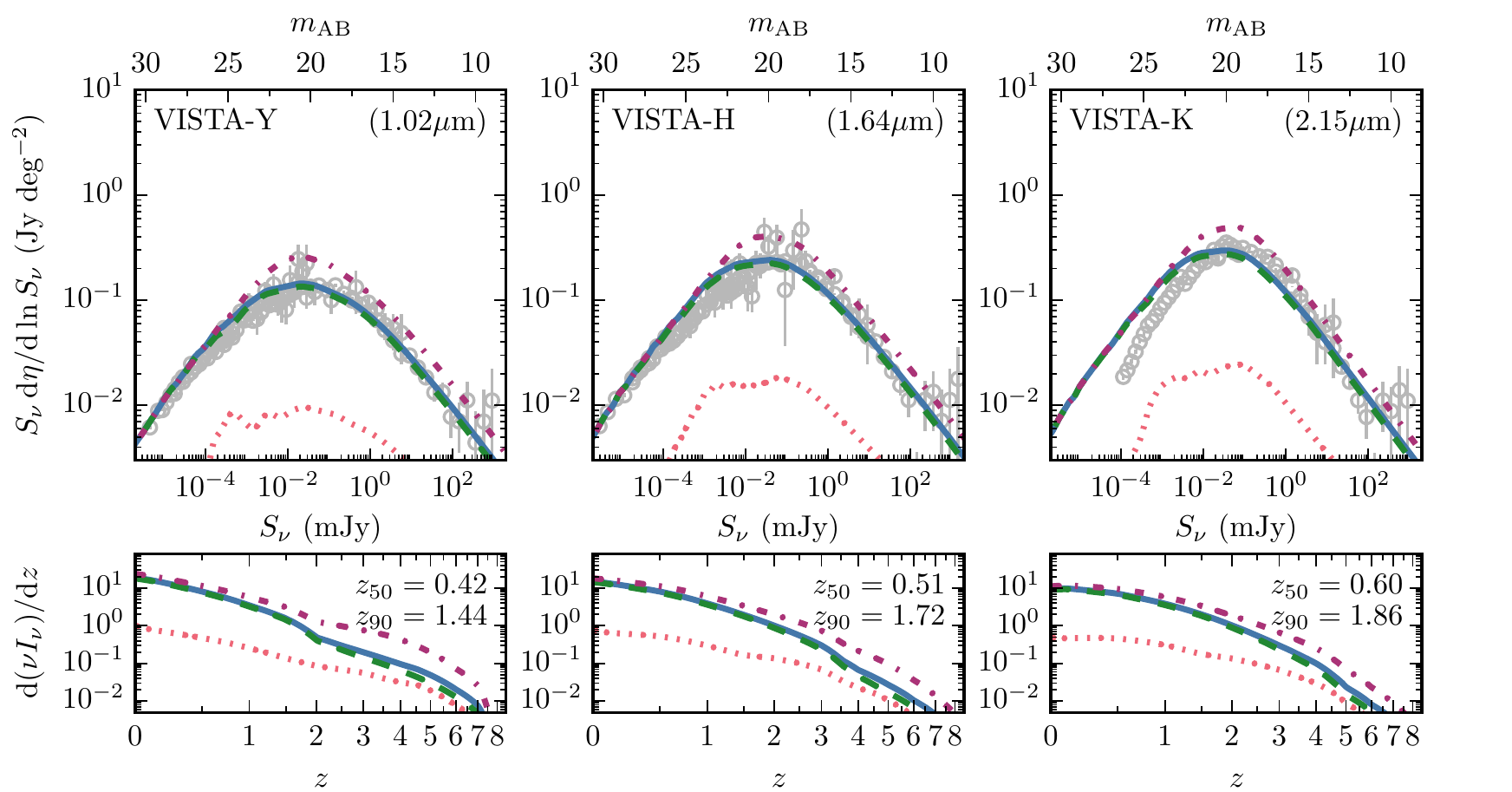}
\caption[]{Continued on the next page.}
\label{fig:xtra_ncts_dndz}
\end{figure*}
\renewcommand{\thefigure}{\thesection\arabic{figure} (Continued)}
\addtocounter{figure}{-1}
\begin{figure*}
\includegraphics[trim = 0 0 0 0, clip = true,width=5.928937in]{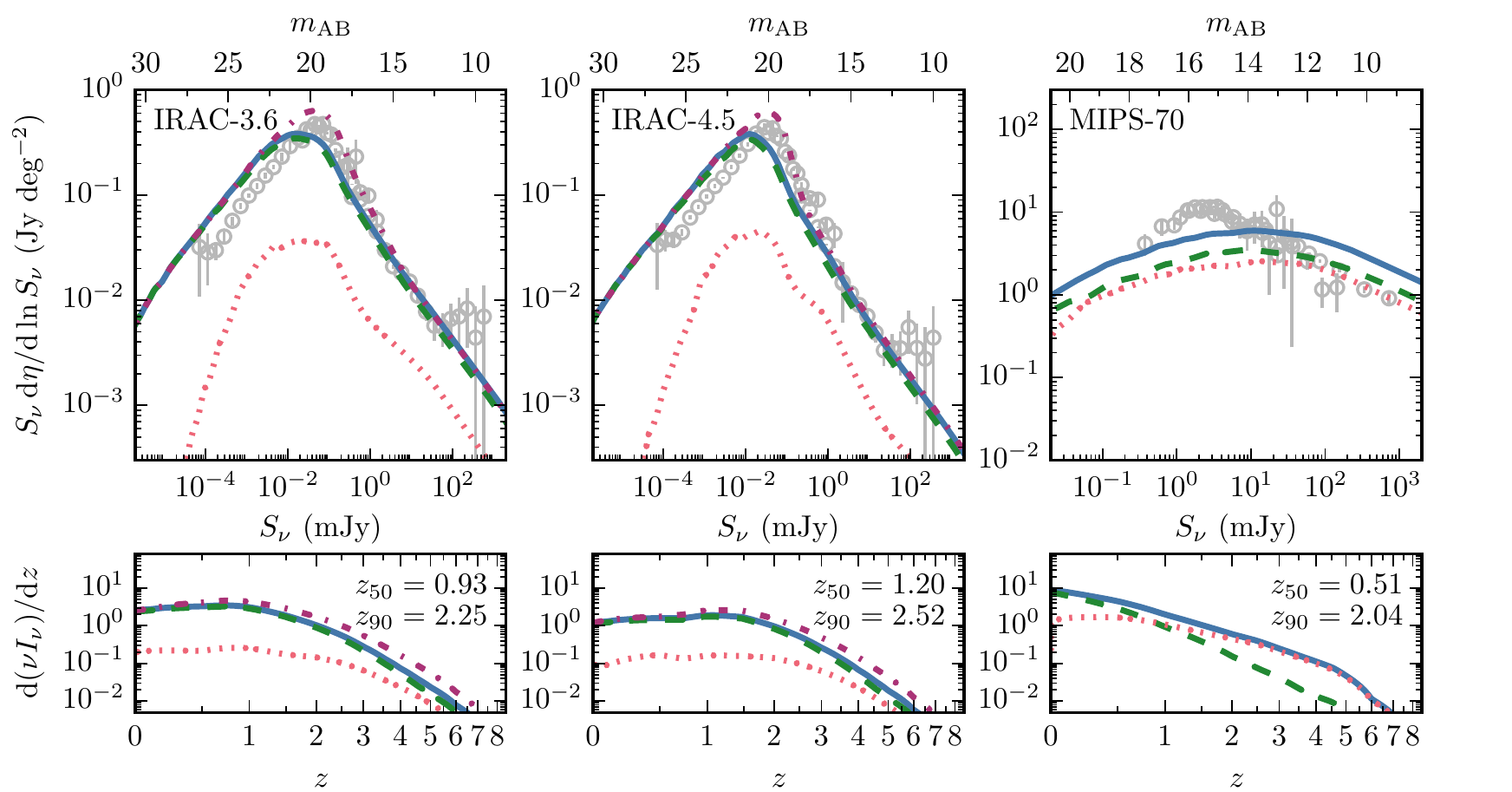}
\includegraphics[trim = 0 0 0 0, clip = true,width=5.928937in]{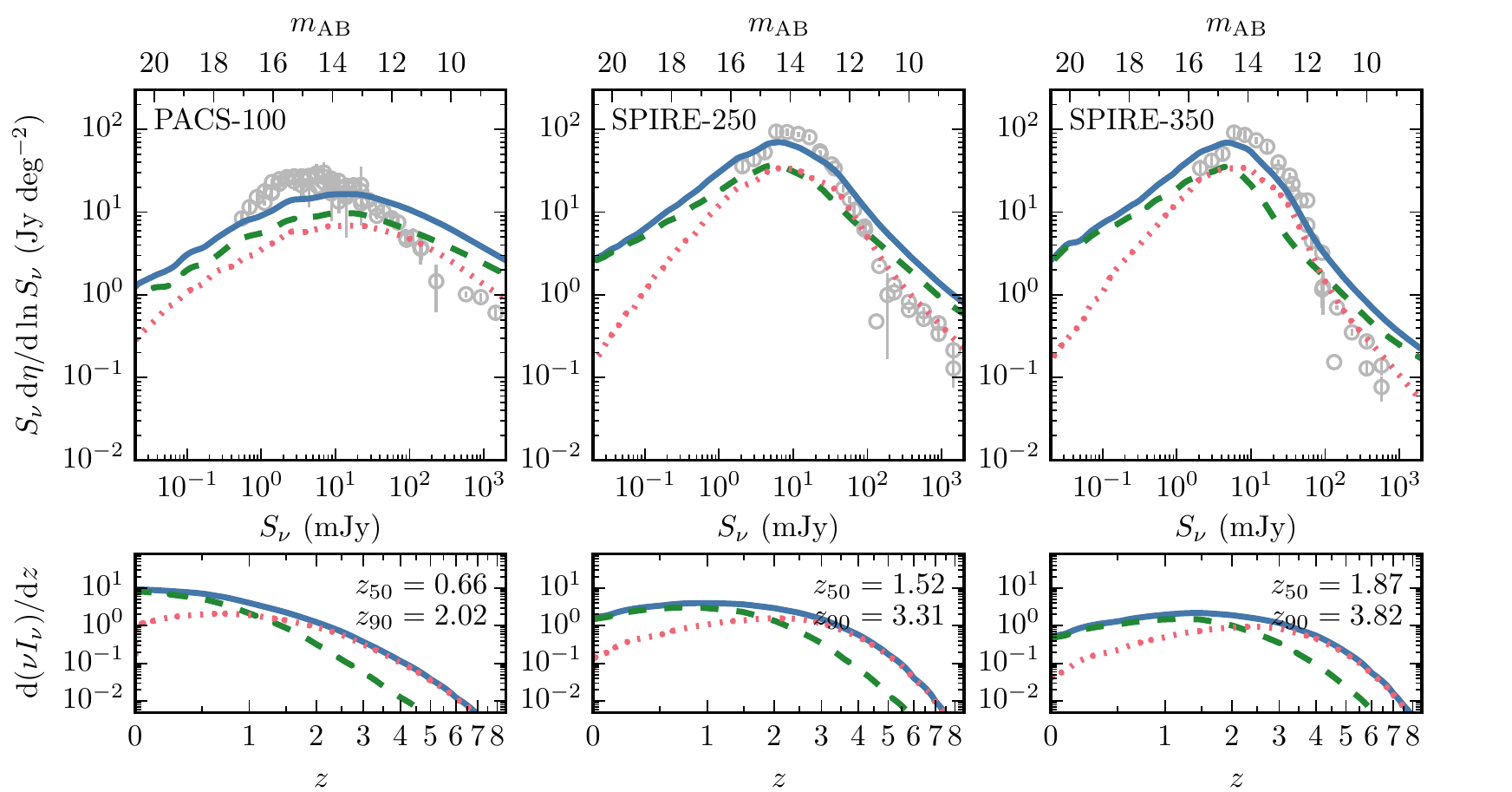}
\caption{Flux-weighted galaxy number counts (major panels) and redshift distribution of background light (in units of nW~m$^{-2}$~sr$^{-1}$) for the band indicated in each panel. Lines have same meaning as in Fig.~\ref{fig:ebl}. Observational data are as compiled by Driver et al. (\citeyear{Driver:2016}). $z_{50}$ and $z_{90}$ correspond to the median and $90^{\mathrm{th}}$ percentile redshifts of the distributions.}
\end{figure*}
\renewcommand{\thefigure}{\thesection\arabic{figure}}

\begin{figure*}
\centering
\includegraphics[trim = 0 0 0 0, clip = true,width=5.928937in]{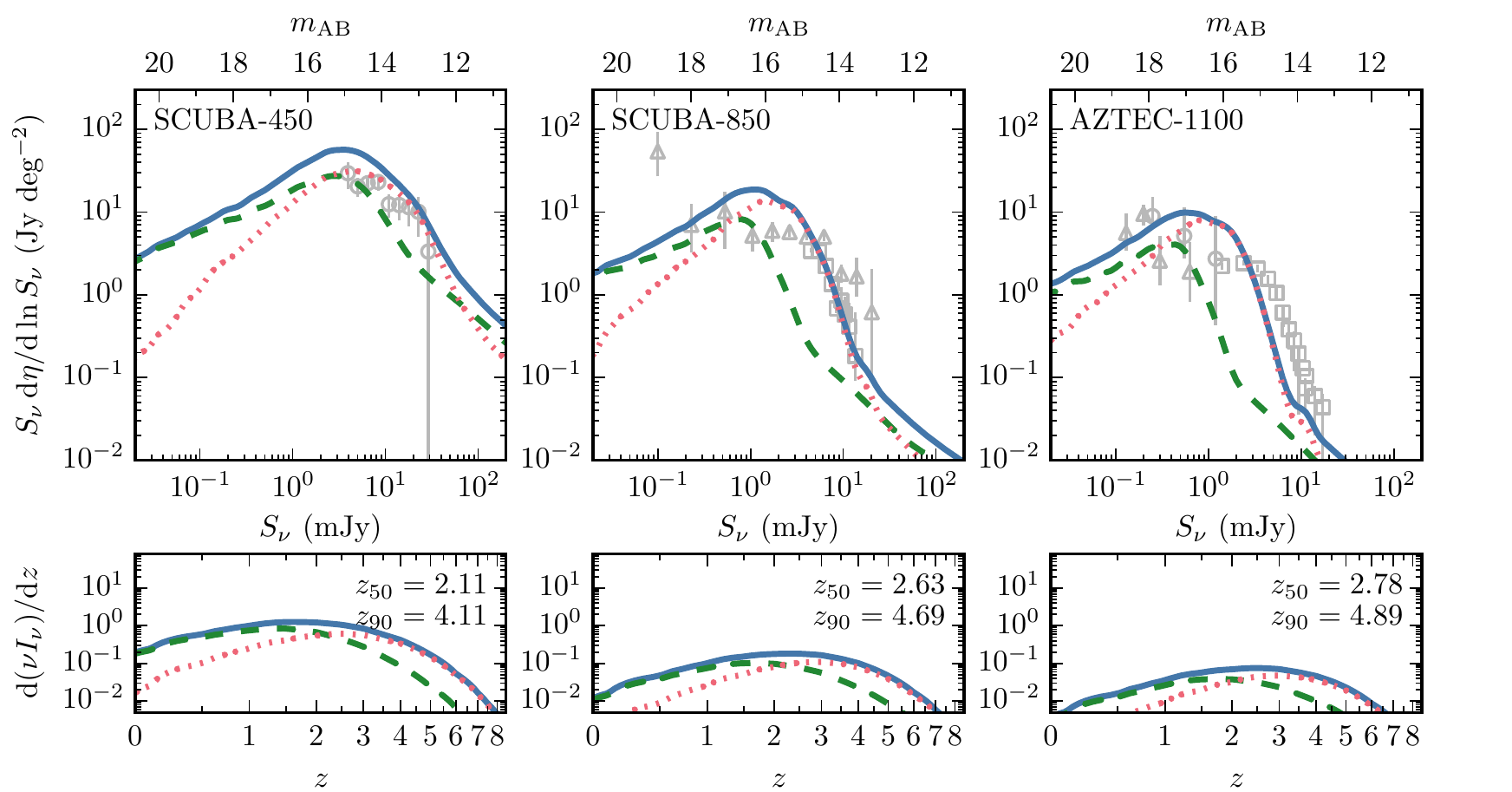}
\caption{Flux-weighted galaxy number counts (major panels) and redshift distribution of background light (in units of nW~m$^{-2}$~sr$^{-1}$) for the band indicated in each panel. Lines have the same meaning as in Fig.~\ref{fig:ebl}. Observational data at (i) $450$~$\muup$m are from Wang et al. (\citeyear{Wang:2017}); (ii) $850$~$\muup$m are from Chen et al. (\citeyear{Chen13}, triangles) and Stach et al. (\citeyear{Stach:2018}, squares); and (iii) $1100$~$\muup$m are from Scott et al. (\citeyear{Scott12}, squares), Carniani et al. (\citeyear{Carniani15}, triangles) and Hatsukade et al. (\citeyear{Hatsukade:2016}, circles). $z_{50}$ and $z_{90}$ correspond to the median and $90^{\mathrm{th}}$ percentile redshifts of the distributions.}
\label{fig:xtra_xtra_ncts_dndz}
\end{figure*}

Fig.~\ref{fig:xtra_ncts_dndz} shows the predicted number counts and redshift distribution of the background light in bands used by Driver et al. (\citeyear{Driver:2016}) but not shown in Fig.~\ref{fig:ncts_dndz}. In Fig.~\ref{fig:xtra_xtra_ncts_dndz} we show these predictions in far-IR bands not considered by Driver et~al. Note that in these figures we do not account for the effect that the coarse angular resolution of single-dish telescopes used for imaging surveys at far-IR wavelengths can have on the derived galaxy number counts \citep[e.g.][]{Karim13}. The impact this issue has on our model predictions is thoroughly explored in \cite{Cowley15}.   

\section{The contribution of optical galaxies to the far-infrared background}
\label{app:optical_dndz}
\begin{figure*}
\centering
\includegraphics[width=6.97522in]{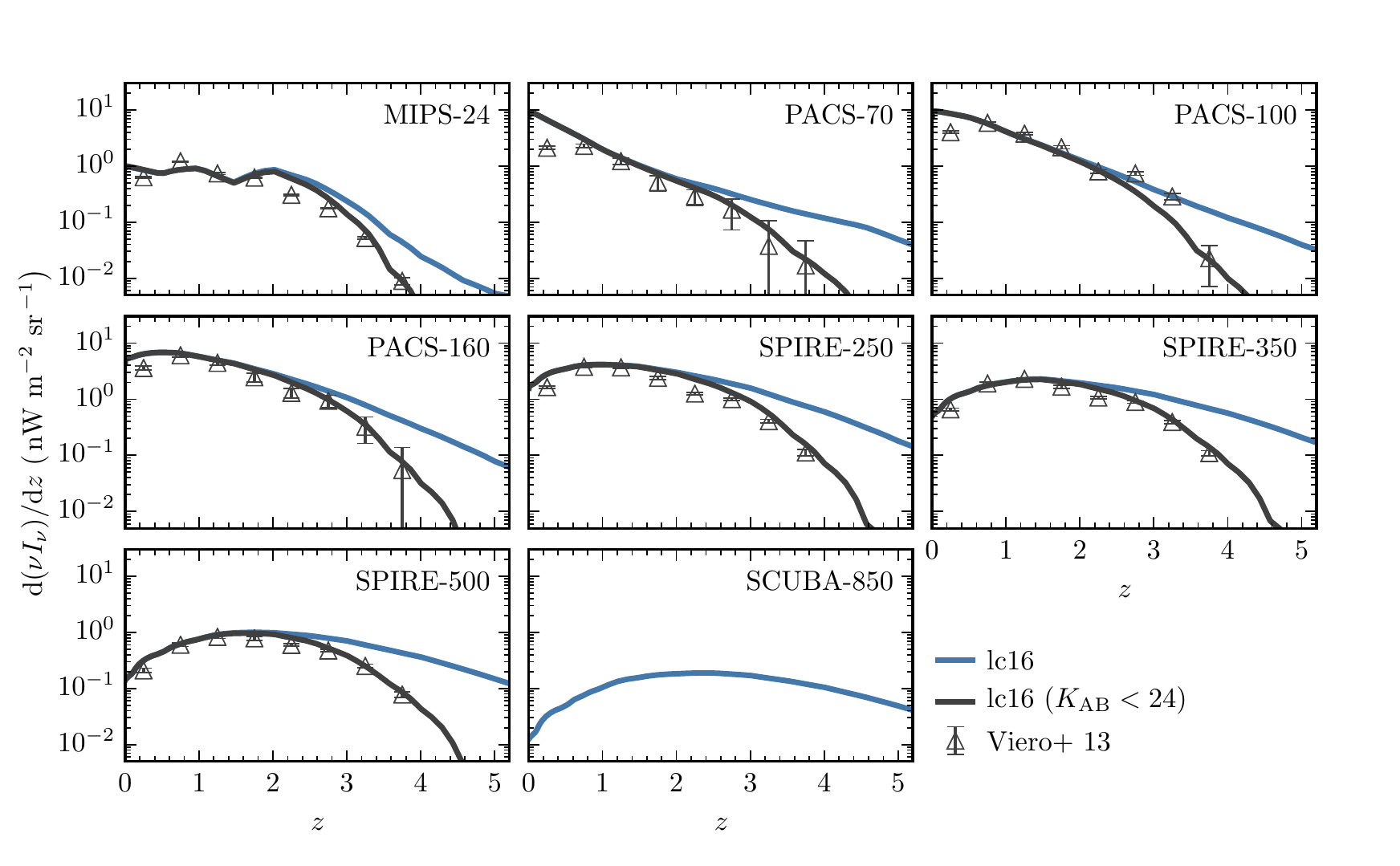}
\caption{As for Fig.~\ref{fig:ebl_dndz_obs}, but the dark grey lines now indicate the predictions for galaxies with $K_{\rm AB}<24$. Observational data are from Viero et al. (\citeyear{Viero:2013b}, open triangles). The estimates of Schmidt et al. have been omitted for clarity.}
\label{fig:optical_dndz}
\end{figure*}
Here we compare our predictions for the distribution of EBL emission redshifts to the observational data of \cite{Viero:2013b}. These authors performed a similar stacking analysis to \cite{Jauzac:2011} and \cite{Bethermin12} on \emph{Herschel} imaging, but began with an input catalogue selected in the near-IR ($K_{\rm AB}<24$), rather than at $24$~$\muup$m. 

We find similarly good agreement with these data as in Fig.~\ref{fig:ebl_dndz_obs}. However, it does appear that the model may overestimate the contribution from the lowest redshift bin, particularly at $70$ and $100$~$\muup$m. This is where the volume probed by Viero et al. will be smallest and combined with their modest $\sim0.63$~deg$^{2}$ area it could be that these lowest-redshift data are not necessarily representative of the Universe. Nevertheless, the overall discrepancy is fairly minor, again showing that the model can plausibly predict the buildup of the EBL at far-IR wavelengths.    

\section{The cosmic star formation history}
\label{app:csfrd}

\begin{figure*}
\centering
\includegraphics[trim = 0 0 0 0, clip = True,width=3.32513in]{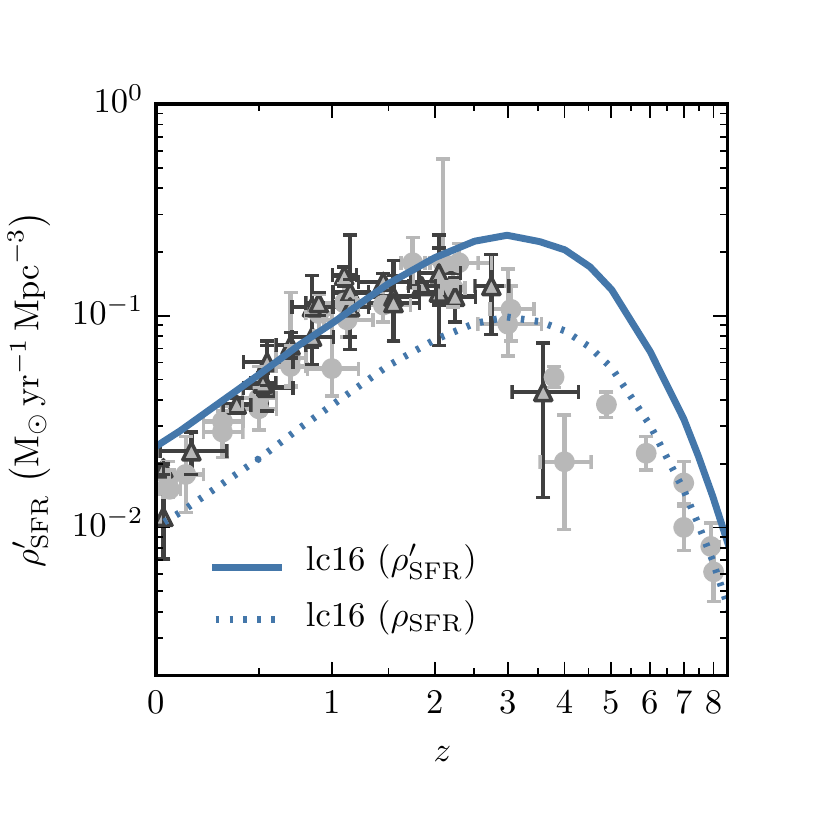}\hspace*{0.32496in}
\includegraphics[trim = 0 0 0 0, clip = True,width=3.32513in]{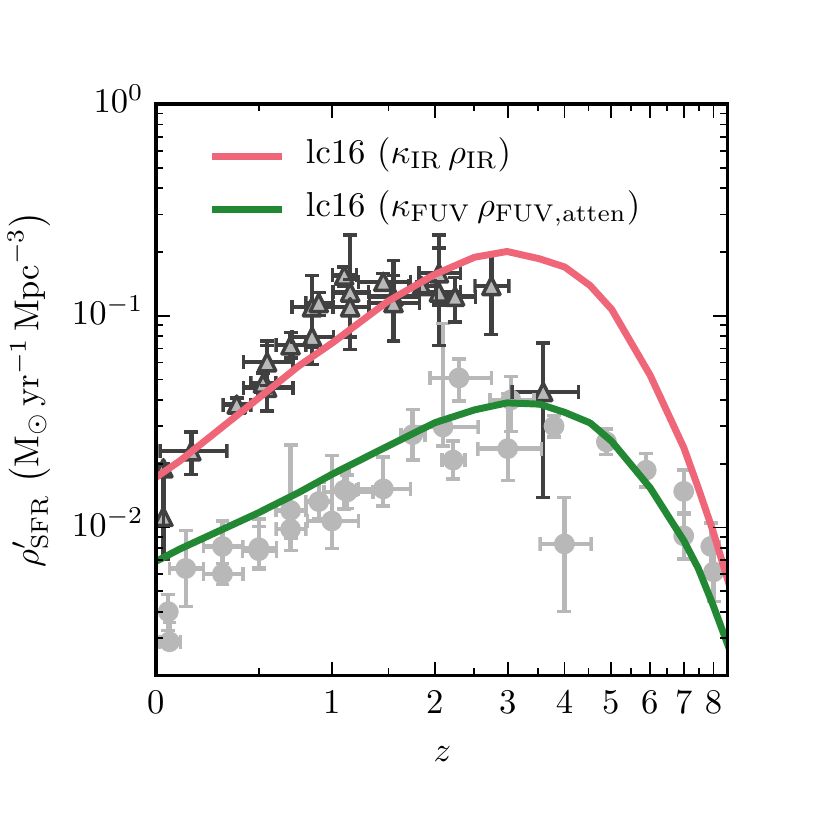}
\caption{The cosmic star formation history. \emph{Left panel}: the difference between the apparent star formation history inferred using Eqn.~\ref{eq:Madau_SFR} (solid line) and the intrinsic prediction of the model (dotted line). \emph{Right panel}: the contribution to the total inferred star formation history from the two terms on the right hand side of Eqn.~\ref{eq:Madau_SFR}, $\kappa_{\rm FUV}\,\rho_{\rm FUV,atten}$ (green line) and $\kappa_{\rm IR}\,\rho_{\rm IR}$ (red line). Observational data in both panels are from the Madau \& Dickinson (\citeyear{MadauDickinson:2014}) compilation, colours and symbols are as in the top-right panel of Fig.~\ref{fig:imf}. In the right panel of this figure we have uncorrected the UV-derived observational estimates for dust attenuation, following the same procedure as Madau \& Dickinson.}
\label{fig:csfrd_app}
\end{figure*}

In the top right panel of Fig.~\ref{fig:imf} we scaled the predicted far-UV and far-IR luminosity densities to apparent star formation rate densities using the same conversion factors as \cite{MadauDickinson:2014} in order to compare our predictions to their compilation of estimates for the cosmic star formation history (see Eqn.~\ref{eq:Madau_SFR}). This is not necessarily the same as the intrinsic cosmic star formation history predicted by the model, $\rho_{\rm SFR}$, and the difference between this and that inferred from Eqn.~\ref{eq:Madau_SFR} is shown in the top panel of  Fig.~\ref{fig:csfrd_app}. The apparent star formation history, $\rho_{\rm SFR}^{\prime}$, is a factor of $\sim2$ greater than $\rho_{\rm SFR}$ for $z\lesssim3$. This factor then decreases towards higher redshifts.

In the left panel of Fig.~\ref{fig:csfrd_app} we show the contribution to the total inferred star formation history from the two terms on the right hand side of Eqn.~\ref{eq:Madau_SFR}. To ease the comparison with the observational data in this panel we have uncorrected the far-UV based observational estimates for dust attenuation, using the same method as Madau \& Dickinson. The attenuated far-UV luminosity density predicted by the model is in reasonable agreement with the observational data over the whole range of redshifts shown. This is unsurprising as this model has been shown to predict evolution of the rest-frame far-UV luminosity function in good agreement with observational estimates, particularly at high redshifts \citep[$6\lesssim z\lesssim 10$,][]{HouJun16,Cowley:2018-JWST}.  The far-IR predictions are in equally reasonable agreement for $z\lesssim3$, beyond which observational estimates become increasingly challenging as discussed earlier. 

Interestingly, the contribution to the apparent star formation history from the far-IR is greater than that from the UV over the entire redshift range shown. This is in contrast to some observational studies \citep[e.g.][]{Bourne:2017,Dunlop:2017}, who argue that the UV contribution dominates for $z\gtrsim4$, though is in agreement with a recent \emph{Herschel} de-blending study (Wang et al. \citeyear{Wang:2019}). This again highlights the need for a consensus regarding the dust-obscured star formation rate density at $z\gtrsim3$.   

\section{The impact of using GRASIL and the recalibration to the P--Millennium}
\label{app:grasil_recal} 
\begin{figure*}
\centering
\includegraphics[trim = 0 0 0 0, clip = True,width=6.97522in]{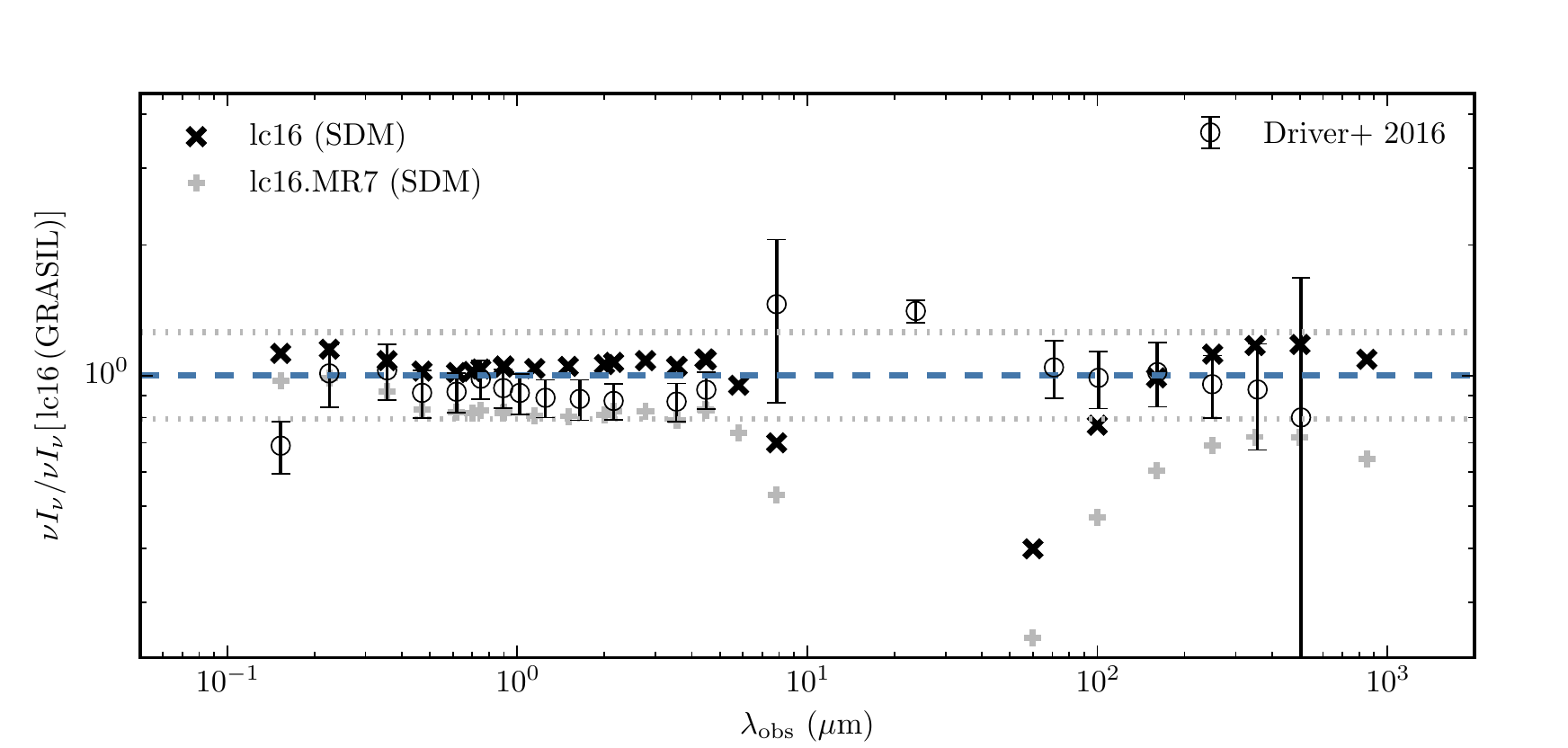}
\caption{The impact of the P--Millennium recalibration and \grasil\ on the predicted EBL. For clarity we have divided everything by the predictions for our fiducial lc16 model with luminosities predicted by \grasil\ i.e. the blue line in Fig.~\ref{fig:ebl}. Predictions from our fiducial lc16 model, and the original model presented in Lacey et al. (\citeyear{Lacey16}) implemented within halo merger trees from the MR7 simulation, both with luminosities predicted by the simple dust model (SDM) described in Lacey et al. (\citeyear{Lacey16}), are shown as black crosses and grey plus signs respectively. Observational data from Driver et al. (\citeyear{Driver:2016}, open circles with errorbars) are shown for reference. A blue dashed line at unity and grey dotted lines at $\pm0.1$~dex are also shown for reference.}
\label{fig:ebl_check}
\end{figure*}
The model used in this work differs slightly from that originally presented by \cite{Lacey16} and in this section we discuss the impact these changes have on the predicted EBL. 

The halo merger trees in which the galaxy formation model is run have been generated from a new dark matter simulation, the P--Millennium \citep{Baugh:2019}, which has different cosmological parameters and a finer halo mass resolution than the MR7 simulation, which was used by Lacey et al; and an improved prescription for the galaxy merger timescale \citep{SimhaCole16} has been implemented. As a result of these changes, Baugh et~al.found it necessary to make small adjustments to two of the galaxy formation parameters such that the original calibration data of Lacey et al. could be reproduced to a similar level of fidelity. As we have stressed earlier, we do not make any further changes to \galform's parameters in this work.

Additionally, here we use \grasil\ for predicting galaxy SEDs (see also \citealt{Cowley:2018-JWST}), whereas in other \galform\ studies the simple dust model (hereafter SDM) described in \cite{Lacey16} is more commonly used.

We assess the impact these two changes (recalibration + \grasil) have on our predictions for the EBL in Fig.~\ref{fig:ebl_check}. For this we compute the EBL for the fiducial model used here (lc16) but with luminosities predicted using the SDM, and for the Lacey et al. model implemented in the MR7 halo merger trees with its original parameter values (lc16.MR7), also with luminosities predicted with the SDM. For ease of highlighting differences between the various models we have divided everything by our predictions for the EBL from our fiducial model with luminosities predicted with \grasil, as presented in Fig.~\ref{fig:ebl}. We can see that for $\lambda_{\rm obs}\lesssim8$~$\muup$m that the lc16 model makes very similar predictions with either the SDM or \grasil, however, there are significant differences at mid-infrared wavelengths that arise from the SDM assumption that the dust is described by only two temperatures (there is an SDM prediction for 24~$\muup$m at $\sim10^{-3}$ that doesn't appear in Fig.~\ref{fig:ebl_check}), with the agreement improving towards far-IR wavelengths. These differences are in line with the comparison of the two dust models performed by \cite{Cowley16SEDs}.

We can also see in Fig.~\ref{fig:ebl_check} that the lc16.MR7 model has less optical/near-IR EBL than lc16 by a factor of $\sim0.1$~dex, with greater differences at longer wavelengths. This indicates that it is the recalibration of the model to the P--Millennium simulation that is responsible for most of the differences between the EBL predictions here and those presented in \cite{Andrews:2018}, who used the model labelled lc16.MR7 (SDM), rather than the use of \grasil, as mentioned earlier. The remaining differences are probably due to approximations made in their numerical integration scheme.

\section{The importance of using a full radiative transfer calculation of the sub-mm flux}
\label{app:grasil_v_h11}

\begin{figure}
    \centering
    \includegraphics[trim = 0 10 0 0, clip = True, width=\linewidth]{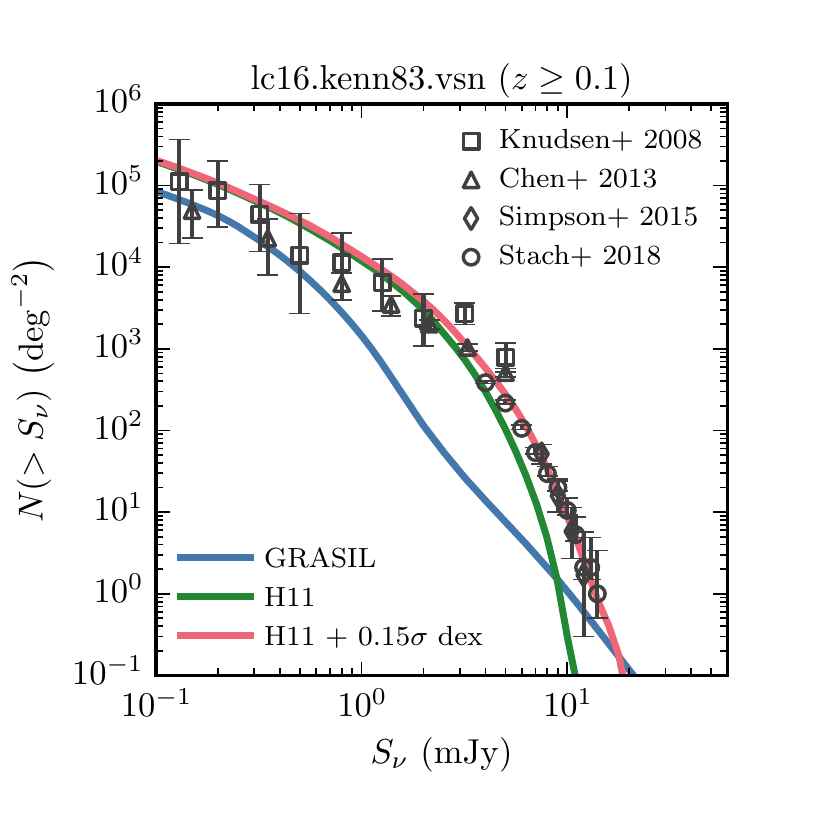}\\
    \includegraphics[trim = 0 10 0 0, clip = True, width=\linewidth]{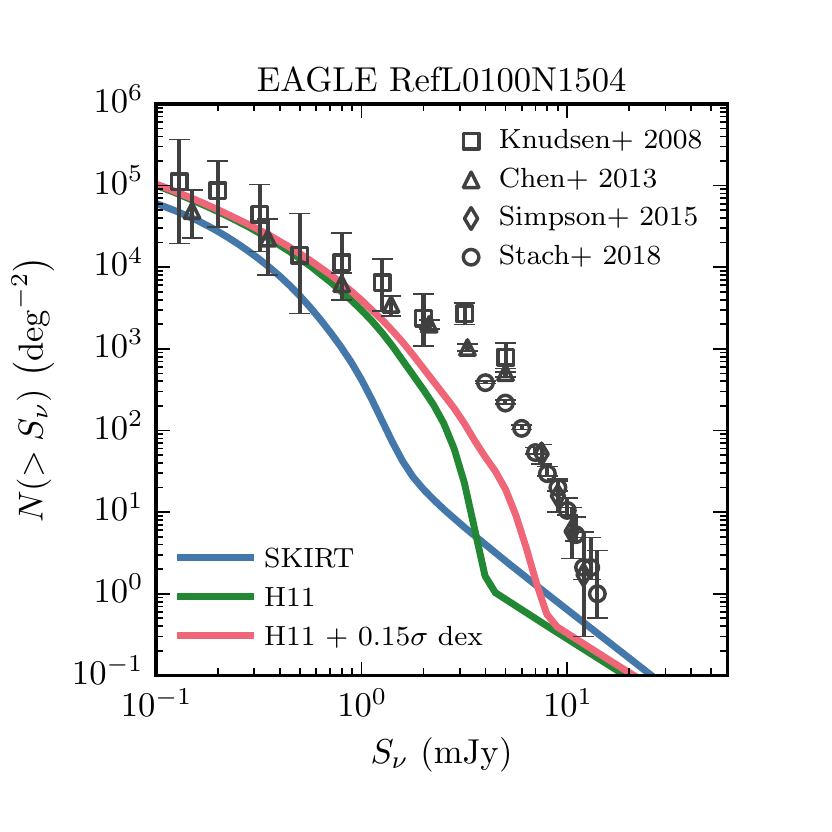}
    \caption{ Predicted galaxy number counts at $850$~$\muup$m. Observational data are as for Fig.~\ref{fig:smg_ncts}. Top panel: predictions for the lc16.kenn83.vsn variant model using {\tt GRASIL} (blue line), the fitting formula proposed by Hayward et al. (\citeyear{Hayward11}, modified to include redshift dependence as  in Safarzadeh et al. \citeyear{Safarzadeh:2017}, green line) and as for the green line but with a further 0.15~dex scatter (red line). Bottom: as for the top panel but using the {\tt EAGLE} simulation predictions as a starting point. We note the visual similarities of these panels to Figure~1 of Safarzadeh et al.}
    \label{fig:ncts_H11}
\end{figure}

Here we use a full radiative transfer calculation to compute SEDs for our simulated galaxies, as this is crucial to make predictions over the full UV-to-mm range that are faithful to the underlying model. However, due to the computational expense and complexity of this calculation, simpler approaches are sometimes preferred. For example, Hayward et al. (\citeyear{Hayward11}, see also Hayward et al. \citeyear{Hayward13a}) proposed a simple approach to calculate the $850 \mu$m flux, using a fitting formula that is based on a small set of idealized hydrodynamical galaxy simulations including radiative transfer. The formula relates the $850 \, \mu $m flux to various intrinsic galaxy properties, such as the star formation rate and dust mass. This relation was subsequently adopted by \cite{Safarzadeh:2017} who used it to claim that the semi-analytical model of \cite{Lu:2014} could reproduce the present-day stellar mass function, cosmic star formation history and sub-mm galaxy number counts whilst assuming a universal solar-neighbourhood IMF, seemingly in contradiction with our conclusions in Section~\ref{sec:top-heavys}. 

Here, we investigate the impact of using a version of this relation, which relates the flux at $850$~$\muup$m to the star formation rate and dust mass in a galaxy (see equation~3 of Safarzadeh et al.). When we apply this fitting formula to our universal IMF variant model (lc16.kenn83.vsn), in a manner completely analogous to that in Safarzadeh et al., we find that this significantly overestimates the resulting $850$~$\muup$m galaxy number counts, relative to the more self-consistent predictions that we obtain with {\tt GRASIL} (see Fig.~\ref{fig:ncts_H11}). We also repeated this exercise using the predictions of the {\tt EAGLE} simulations as a starting point, finding a similar discrepancy when compared to {\tt EAGLE} collaboration predictions derived from the {\tt SKIRT} radiative ransfer code (also shown in Fig.~\ref{fig:ncts_H11}). 

Our conclusion is therefore not that the Lu et al. model is able to reproduce the sub-mm number counts using a universal solar-neighbourhood IMF, but rather that the use of a simple fitting formula for the sub-mm flux introduces a significant systematic error into the predictions, making this an inappropriate substitute for a full radiative transfer calculation. We therefore advise extreme caution regarding the use of simple fitting formulae for predicting dust emission in cosmological-scale galaxy formation simulations.
     
%%%%%%%%%%%%%%%%%%%%%%%%%%%%%%%%%%%%%%%%%%%%%%%%%%
% Don't change these lines

\bsp	% typesetting comment
\label{lastpage}
\end{document}